\documentclass[acmtog, authorversion]{acmart}

\newcommand{\final}{1}

\usepackage[english]{babel}
\usepackage{ifthen}

\usepackage{amsmath}
\usepackage{bm}        % Bold math symbols
\usepackage{textcomp}  % Registered and Copyright symbols
\usepackage{mathtools} % Enable \coloneqq

\usepackage[toc,page,titletoc]{appendix}
\usepackage[capitalize,nameinlink]{cleveref}
\crefname{supp}{Supplement}{Supplements}

\usepackage{caption}
\usepackage{subfig}
\usepackage[export]{adjustbox}

\usepackage{xspace}

\usepackage{enumitem}

\usepackage{array}

\providecommand\BibTeX{{\normalfont B{\scshape i\kern-0.0em b}\TeX}}

\newcommand{\warning}[1]{{\it\color{red} #1}}
\newcommand{\note}[1]{{\it\color{blue} #1}}
\definecolor{SithColor}{rgb}{0.8,0.5,0} % color for Sith
\newcommand{\qisun}[1]{{\color{SithColor} Qi: #1 $\qed$}}
\definecolor{ConsularColor}{rgb}{0,0.7,0} % color for the Jedi Consulars (e.g. Yoda)
\newcommand{\anjul}[1]{{\color{ConsularColor} Anjul: #1 $\qed$}}
\definecolor{GuardianColor}{rgb}{0,0,0.8} % color for the Jedi Guardians (e.g. Obiwan)
\newcommand{\monde}[1]{{\color{GuardianColor} Monde: #1 $\qed$}}
\newcommand{\ben}[1]{{\color{ConsularColor} Ben: #1 $\qed$}}
\newcommand{\rachel}[1]{{\color{purple} Rachel: #1 $\qed$}}

\newcommand{\nothing}[1]{}

\ifthenelse{\equal{\final}{1}} {
    \renewcommand{\warning}[1]{} 
    \renewcommand{\note}[1]{}
    \renewcommand{\qisun}[1]{} 
    \renewcommand{\ben}[1]{}
    \renewcommand{\monde}[1]{}
    \renewcommand{\rachel}[1]{}
    \renewcommand{\anjul}[1]{}
}

\newcommand{\Caption}[2]{\caption[#1]{{\em #1} #2}}

\DeclareMathOperator\erfc{erfc}

\DeclareMathOperator*{\argmin}{arg\,min}

\newcommand{\minDepth}{d_\textit{min}}

\newcommand{\ipd}{w_\textit{IPD}}

\newcommand{\leye}{\textit l}
\newcommand{\reye}{\textit r}
\newcommand{\vergent}{\textit v}
\newcommand{\saccade}{\textit s}

\newcommand{\ang}{\alpha}

\newcommand{\vergentAng}{\ang_\vergent}
\newcommand{\saccadeAng}{\ang_\saccade}
\newcommand{\combineAng}{\bm\ang}
\newcommand{\leftAng}{\ang_\leye}
\newcommand{\rightAng}{\ang_\reye}

\newcommand{\vergentAmp}{\Delta\vergentAng}
\newcommand{\saccadeAmp}{\Delta\saccadeAng}
\newcommand{\combineAmp}{\Delta\bm\ang}

\newcommand{\rbfWeight}{w}
\newcommand{\rbfCenter}{\mathbf{c}}

\newcommand{\travelTime}{\mathcal T}

\newcommand{\mean}{\mu}
\newcommand{\std}{\sigma}
\newcommand{\decay}{\tau}

\newcommand{\expectation}{\mathbb{E}}

\newcommand{\conditionEvalAccuracy}{\mathbf{C}}
\newcommand{\conditionAccuracyFull}{\mathbf{FULL}}
\newcommand{\conditionAccuracyVergence}{\mathbf{VER}}
\newcommand{\conditionAccuracySaccade}{\mathbf{SAC}}

\newcommand{\conditionEvalStudyVergent}{\mathbf{C_s}}
\newcommand{\conditionEvalStudyCombinedShort}{\mathbf{C_m}}
\newcommand{\conditionEvalStudyCombinedLong}{\mathbf{C_l}}

\newcommand{\PDF}{f}
\newcommand{\timestamp}{t}

\newcommand{\loss}{L}

\newcommand{\vergenceAdj}{vergence\xspace}
\newcommand{\saccadeAdj}{saccadic\xspace}
\newcommand{\KLdiv}{KLdiv\xspace}

\graphicspath{
{figures/}
}

\setcopyright{acmlicensed}
\acmJournal{TOG}
\acmYear{2023}
\acmVolume{42}
\acmNumber{6}
\acmArticle{222}
\acmMonth{12}
\acmPrice{15.00}
\acmDOI{10.1145/3618334}

\acmSubmissionID{282}
\begin{document}

%%%%%
%% space saver hack
%%%%%
\setlength{\abovecaptionskip}{1.0ex}
\setlength{\belowcaptionskip}{1.0ex}
\setlength{\floatsep}{1.0ex}
\setlength{\dblfloatsep}{\floatsep}
\setlength{\textfloatsep}{2.0ex}
\setlength{\dbltextfloatsep}{\textfloatsep}
\setlength{\abovedisplayskip}{1.0ex}
\setlength{\belowdisplayskip}{1.0ex}
\Urlmuskip=0mu  plus 10mu

%% The "title" command has an optional parameter,
%% allowing the author to define a "short title" to be used in page headers.
%% \title[short title]{full title}
%% Capitalize your title: https://capitalizemytitle.com/
%\title[Towards Optimizing Visual Performance with Immersive Displays]{Towards Optimizing Visual Performance with Immersive Displays:\\ Modeling the Vergence-Saccade Temporal Interference}

\title[Probability-Modeled Stereoscopic Eye Movement Completion Time in VR]{The Shortest Route Is Not Always the Fastest:\break Probability-Modeled Stereoscopic Eye Movement Completion Time in VR}

%% The "author" command and its associated commands are used to define
%% the authors and their affiliations.
%% Of note is the shared affiliation of the first two authors, and the
%% "authornote" and "authornotemark" commands
%% used to denote shared contribution to the research.
\author{Budmonde Duinkharjav}
\affiliation{%
  \institution{New York University}
  \country{USA}}
\email{budmonde@gmail.com}

\author{Benjamin Liang}
\affiliation{%
  \institution{New York University}
  \country{USA}}
\email{ben.liang@nyu.edu}

\author{Anjul Patney}
\affiliation{%
  \institution{NVIDIA}
  \country{USA}}
\email{anjul.patney@gmail.com}

\author{Rachel Brown}
\affiliation{%
  \institution{NVIDIA}
  \country{USA}}
\email{rachelabrown347@gmail.com}

\author{Qi Sun}
\affiliation{%
  \institution{New York University}
  \country{USA}}
\email{qisun0@gmail.com}
%% By default, the full list of authors will be used in the page
%% headers. Often, this list is too long, and will overlap
%% other information printed in the page headers. This command allows
%% the author to define a more concise list
%% of authors' names for this purpose.
%\renewcommand{\shortauthors}{Anonymous}

%%
%% The code below is generated by the tool at http://dl.acm.org/ccs.cfm.
%% Please copy and paste the code instead of the example below.
%%


%%
%% Keywords. The author(s) should pick words that accurately describe
%% the work being presented. Separate the keywords with commas.

\begin{teaserfigure}
    \centering
    \includegraphics[width=\linewidth]{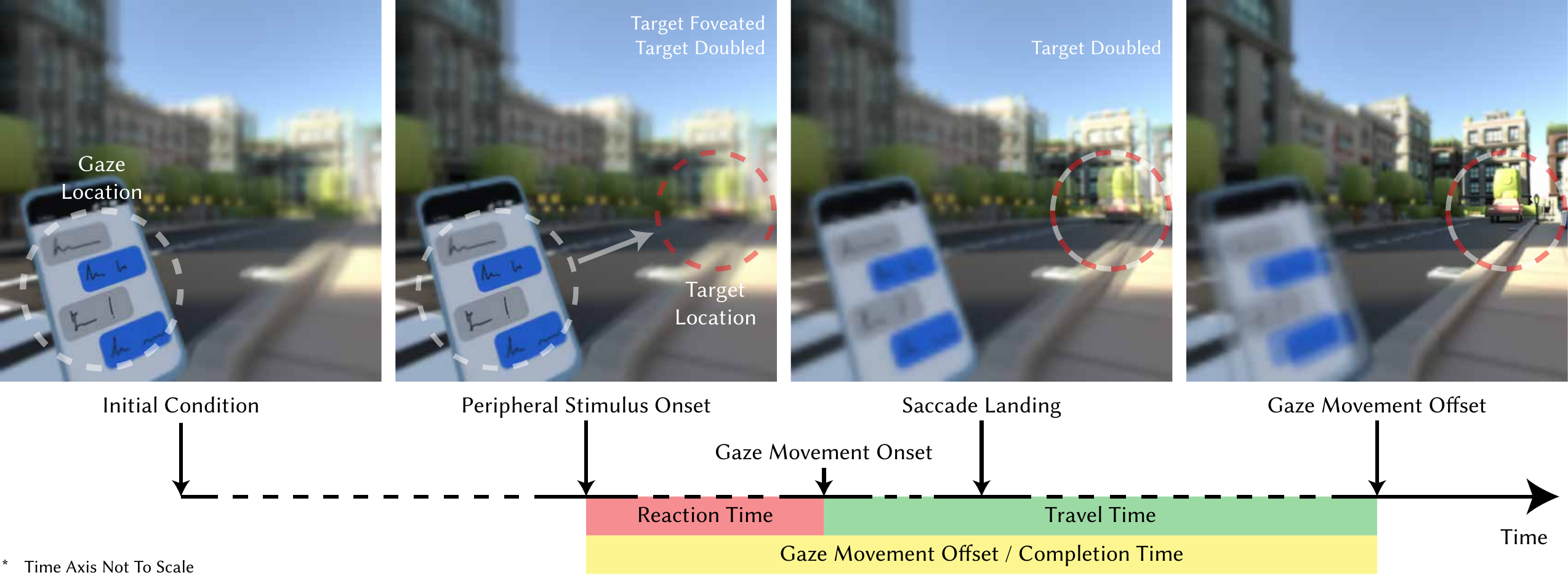}
  \Caption{
    Predicting stereoscopic eye movement completion time.
  }{%
  Our model predicts the completion time of stereoscopic eye movement toward a target in the visual field.
  It provides a probability distribution of the duration between the onset of the target and the first moment we can see it in a clear, unfoveated manner, accounting for both saccadic and vergence changes necessary to do so.
  (a) When users focus their gaze on a specific 3D target, objects at a different depth appear doubled (due to incorrect vergence), blurred (due to peripheral vision), or both.
  (b) A peripheral target might trigger a gaze movement of both the left and right eyes, which initiates after a reaction time necessary for cognitive processing.
  (c) First, the conjugate and ballistic saccadic movement lands the target into foveal vision with high acuity;
  but we may still see double vision due to the slower, incomplete vergence movement.
  (d) Once the disconjugate vergence movement also completes, we can successfully fuse the stereoscopic imagery and see the target clearly.
  }
  \label{fig:teaser}
\end{teaserfigure}

\begin{abstract}
Speed and consistency of target-shifting play a crucial role in human ability to perform complex tasks.
Shifting our gaze between objects of interest quickly and consistently requires changes both in depth and direction.
Gaze changes in depth are driven by slow, inconsistent \emph{vergence movements} which rotate the eyes in opposite directions, while changes in direction are driven by ballistic, consistent movements called \emph{saccades}, which rotate the eyes in the same direction.
In the natural world, most of our eye movements are a combination of both types.
While scientific consensus on the nature of saccades exists, vergence and combined movements remain less understood and agreed upon.

We eschew the lack of scientific consensus in favor of proposing an operationalized computational model which predicts the completion time of any type of gaze movement during target-shifting in 3D.
To this end, we conduct a psychophysical study in a stereo VR environment to collect more than 12,000 gaze movement trials, analyze the temporal distribution of the observed gaze movements, and fit a probabilistic model to the data.
We perform a series of objective measurements and user studies to validate the model.
The results demonstrate its predictive accuracy, generalization, as well as applications for optimizing visual performance by altering content placement.
Lastly, we leverage the model to measure differences in human target-changing time relative to the natural world, as well as suggest scene-aware projection depth.
By incorporating the complexities and randomness of human oculomotor control, we hope this research will support new behavior-aware metrics for VR/AR display design, interface layout, and gaze-contingent rendering.
\end{abstract}

\maketitle

\ifthenelse{\equal{\final}{0}}
{\input{sec_timeline}}
{}

\section{Introduction}
\label{sec:introduction}
Gaze movement patterns are dictated by the strengths and limitations of the visual system.
Visual acuity is much higher in the central region of the retina, encouraging observers to first shift their gaze to bring targets of interest into the fovea prior to analyzing any details.
Furthermore, the binocular nature of human vision dictates that both left and right eyes must move in coordination to focus at the same location.
Consequently, several distinct classes of eye movement patterns have evolved in humans to fulfill various roles and are used in different situations.
Due to the underlying neurological and mechanical limitations of eye movements, each one exhibits distinct performance characteristics;
some are slow and steady, while others are ballistic and jerky.
The combination of all classes of movements forms an efficient and comprehensive overall gaze behavior strategy in 3D visual environments.

The speed of these movements are critical in complex tasks such as driving, where we rapidly move our eyes to acquire a plethora of information from the surroundings such as the presence of pedestrians, the approaching of vehicles, the speedometer reading, and even GPS navigation instructions.
In those tasks, there is always a delay between the decision to acquire a visual target, and our two eyes successfully landing on it.
We ask ``how long is this delay and how does it depend on the displacement of our gaze location?''.
With the emerging adoption of virtual/augmented reality (VR/AR), answering this question enables us to design 3D content that allows for an efficient target changing.

\begin{figure*}
  \centering
  \subfloat[definition of measured angles]{
      \includegraphics[width=.23\linewidth]{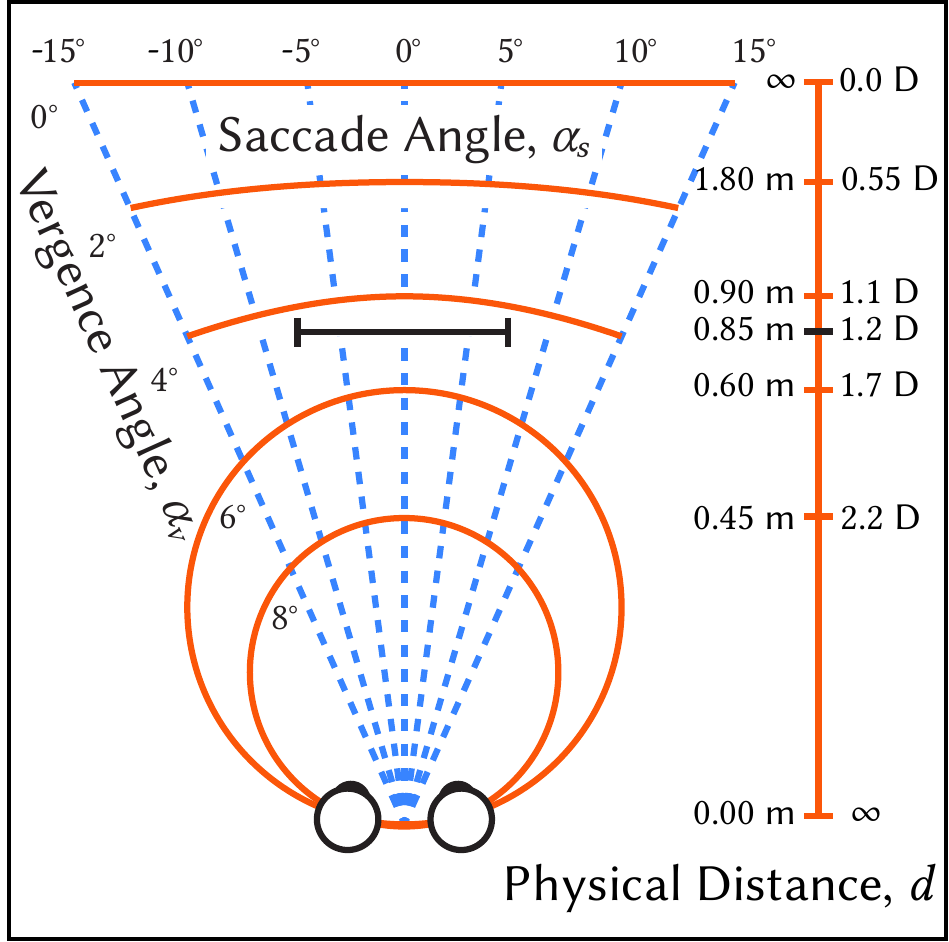}
      \label{fig:background:angles}
  }%subfloat
  \hfill
  \subfloat[symmetrical vergence motion]{
      \includegraphics[width=.23\linewidth]{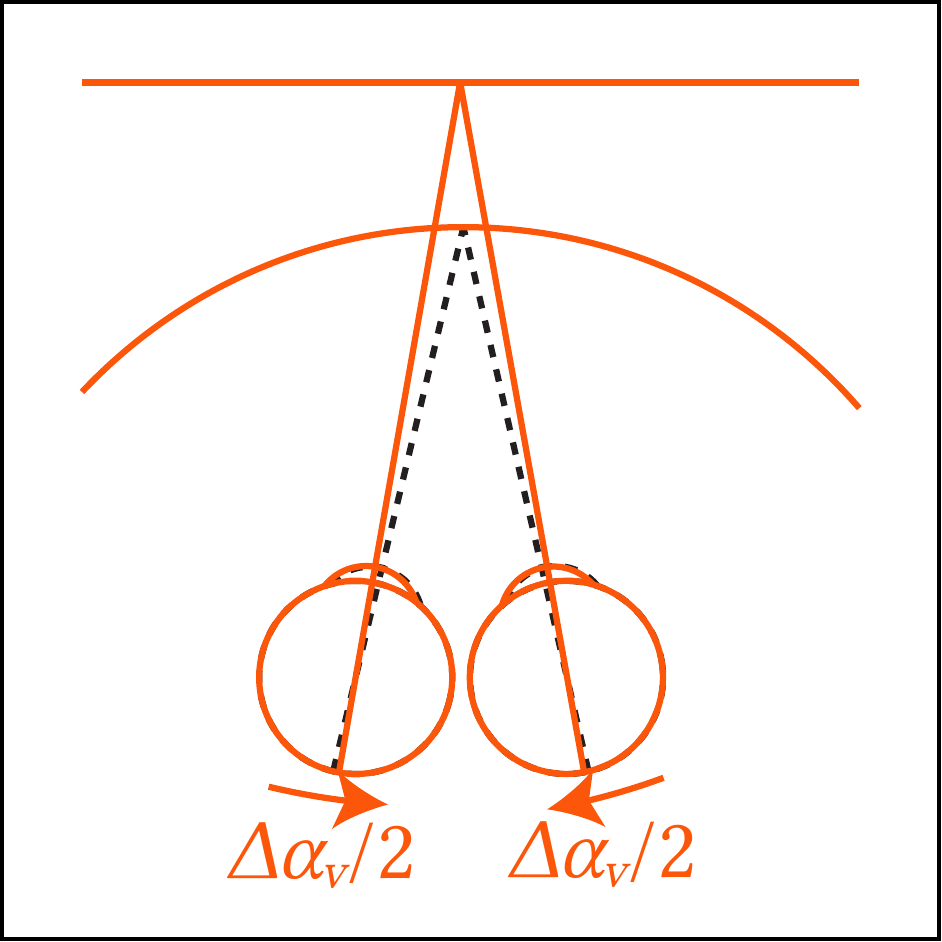}
      \label{fig:background:vergence}
  }%subfloat
  \hfill
  \subfloat[conjugate saccade motion]{
      \includegraphics[width=.23\linewidth]{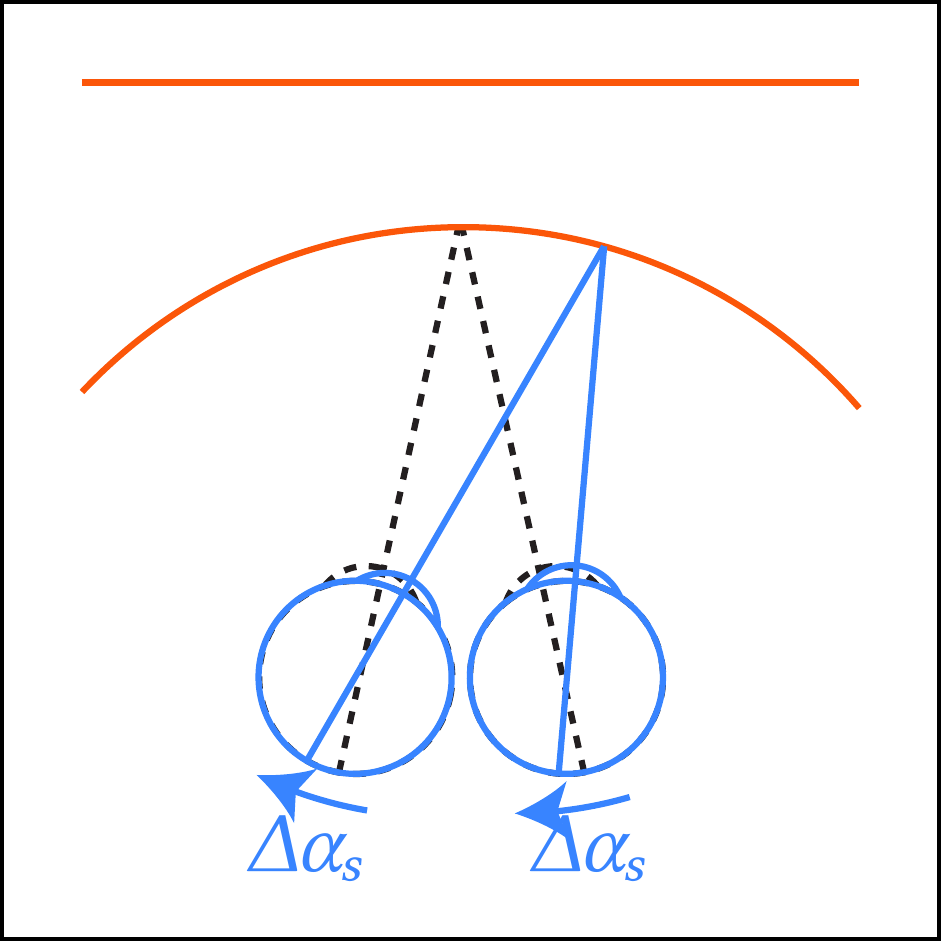}
      \label{fig:background:saccade}
  }%subfloat
  \hfill
  \subfloat[disconjugate combined motion]{
      \includegraphics[width=.23\linewidth]{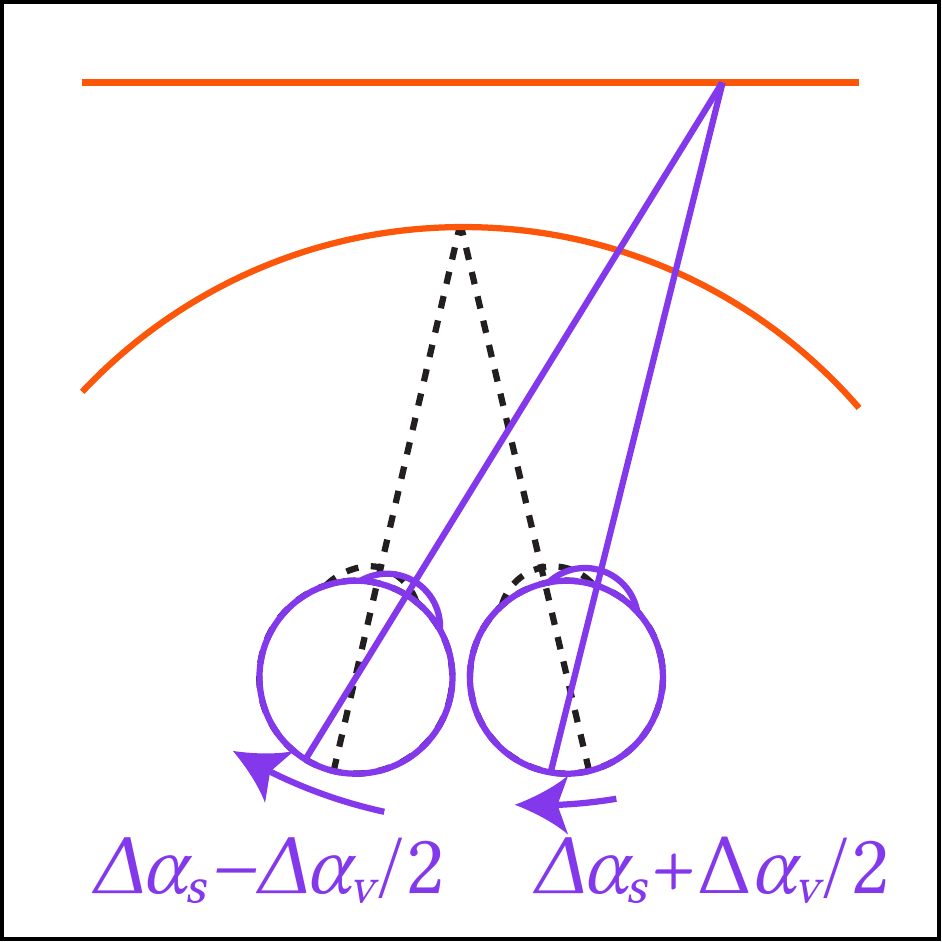}
      \label{fig:background:combined}
  }%subfloat
  \Caption{
      Illustration of various eye movements. 
  }{%
      \subref{fig:background:angles} We illustrate how we define and measure the angles of eye \vergenceAdj movements $\vergentAng$, and \saccadeAdj movements $\saccadeAng$ throughout the paper.
      For further intuition, the physical distance of objects appearing at $\saccadeAng = 0^\circ$ is illustrated in units of meters, and Diopters (i.e., reciprocal of meters).
      Here, interpupillary distance (IPD) is chosen to be equal to the human average of $63\ \text{mm}$ \cite{fesharaki2012ipd}.
      The optical display depth of the headset is overlaid as a horizontal black bar at a depth of $0.85\ \text{m}$, or $1.2\ \text{D}$.
      \subref{fig:background:vergence} In \vergenceAdj motion, the two eyes move symmetrically in opposing directions; away from each other in divergent movement and towards each other in convergent movement.
      \subref{fig:background:saccade} In \saccadeAdj motion, both eyes rotate by the same amount in the same direction.
      \subref{fig:background:combined} In combined motion, each eye moves a different amount. The rotation of each eye can be derived as the sum and difference of the corresponding \vergenceAdj and \saccadeAdj coordinate shift as defined in \subref{fig:background:angles}.
  }
  \Description{Illustration of various eye movements.}
  \label{fig:background}
\end{figure*}

Prior vision science studies suggest that gaze shifts move along two primary axes (\Cref{fig:background:angles}): one in \emph{direction} and the other in \emph{depth} \cite{zee1992saccade}.
Highly rapid and consistent eye motions that quickly shift to a peripheral location, called \emph{saccades}, are crucial for fast reaction to targets in different directions.
In contrast, eye movements that shift the gaze in depth by rotating each eye in opposing directions, called \emph{vergence movements}, are relatively slower and more inconsistent.
Often, both of these movements are executed concurrently, and the performance of such \emph{combined} movements exhibit a different time signature which is faster than pure vergence movements, but slower than pure saccades \cite{zee1992saccade,bucci2006latency,lang2014saccade,yang2004saccade}.
While vision science literature has extensively studied saccadic movements and provided comprehensive models for its temporal characteristics (i.e., the main sequence \cite{bahill1975main,van2008saccadic}), the nature of vergence and combined movements exhibit confounding theories \cite{chen2010behaviors,cullen2011neural,king2011binocular}.

As an alternative, we present the first operational model that predicts the required eye movement completion time necessary for shifting the gaze to new 3D targets in stereoscopic virtual environments.
We recognize the current lack of first-principle consensus on how vergence/combined eye movements are neurologically constructed. 
Additionally, we note that noise in both human behavior and eye-tracking adds difficulty to comprehensive study of complex stereoscopic movements with downstream applications.
Circumventing these obstacles, we take a holistic approach to (1) focus on \emph{when} both eyes land on a target after its onset, instead of the intermediate trajectory;
and (2) form a computational model which accounts for the noise and variability to produce a \emph{probabilitic} prediction, instead of a deterministic one.

We fit our model and validate its accuracy using our psychophysical study data, which includes more than $12,000$ individual trials to measure the temporal offsets of gaze movements in a stereo VR environment.
The results evidence the model's consistent prediction accuracy, generalizability to unseen participants and trials, as well as the capability of forecasting and optimizing task performance with various real-world VR scenarios.
Our model can be applied to measure the difficulty of video games in VR and how the scale of variability in depth can alter gaze movement behaviors for users.
We also explore how completion time predictions can be used as a metric for evaluating the placement of 3D UI elements in VR/AR applications.
Recalling the driving example, we can improve driver awareness by placing a virtual car dashboard overlay (with speedometer readings and navigation instructions etc.) in an adaptive manner to minimize completion times of objects that appear in the driver's periphery in changing surrounding environments.

This research aims to propose an operational model for computer graphics applications for a behavioral phenomenon that is yet to be fully understood.
We believe that providing a quantitative understanding of how emerging VR/AR technology influences statistical signatures of human target-changing performance during daily tasks is beneficial even without the neurological understanding of the underlying behaviors.
We hope the research can serve as a novel benchmark to guide 3D interfaces and act as a metric for the ``user performance'' in various applications and mediums. To this aim, we will release the source code and de-identified study data at \url{www.github.com/NYU-ICL/stereo-latency}.
In summary, our main contributions include:
\begin{itemize}[leftmargin=*]
    \item a series of psychophysical studies and data which systematically characterize visual performance (measured by completion/offset time) across various vergence-saccade combined eye movements in VR;
    \item an operational model that predicts the statistical distribution of completion times;
    \item demonstration of the model's accuracy and effectiveness in predicting and optimizing VR users' target-changing performance in natural scenarios;
    \item model application to measure users' visual performance discrepancies among various games, 2D and VR displays, as well as recommendations for depth designs for 3D user interfaces.
\end{itemize}

\section{Related Work}
\label{sec:related}

\subsection{Eye Movement, Visual Behaviors, and Performance}
Human eyes are highly dynamic, consisting of various types of movements including smooth pursuit, vestibulo-ocular, saccade, and vergence movements.
Saccade and vergence are the two most frequent movements to redirect gaze in 3D spaces \cite{lang2014saccade}.
There has been extensive study of them in the context of computer graphics, displays, and interactions \cite{yarbus2013eye,hadnett2019effect}. 
Unlike most traditional desktop displays, VR/AR platforms provide high field-of-view
stereoscopic displays, which simultaneously unlock both saccade and vergence movements.
Understanding the timing of these visual movements is essential in broad applications such as esports \cite{Duinkharjav:2022:IFI}, driving \cite{salvucci2002time}, and healthcare \cite{bertram2016eye}. 

Pure saccades are rapid and conjugate eye movements that change the direction of gaze along a circle of iso-vergence (or the geometric horopter) which is computed using the centers of the two eyes and the fixation point (\Cref{fig:background}).
In the scope of this work, we simplify the measurements by equalizing the optical and visual axes (cf. \cite{konrad2020gaze,krajancich2020optimizing}), leaving the study of this difference as future work.
Saccades are high-speed, ballistic motions with short travel times and a probability distribution of spatial error skewing towards undershooting the target location \cite{lisi2019gain}.
The scan path, speed, and spatial accuracy of a saccade are all influenced by the characteristics of the visual content \cite{sitzmann2018saliency,martin2022scangan360,Duinkharjav:2022:IFI,van2007sources,specht2017simple,arabadzhiyska2017saccade}, and have been extensively studied and modeled  \cite{bahill1975main,boghen1974velocity,van2008saccadic}.
Although those features can also be influenced by visual tasks \cite{hu2021fixationnet,hu2021ehtask}, studies on the \emph{main sequence} \cite{bahill1975main} show the consistency in completion time after the ocular-motor-controlled movement starts, independent of cognitive factors.

By comparison, pure vergences are both slower and disconjugate, directing the gaze to a new location in depth and thereby defining a new geometric horopter.
In stereo displays that lack accommodative cues, the displacement of the images presented to the two eyes provides an essential depth cue that drives vergence eye movements.
In the context of VR/AR, the conflict between the variable vergence cues provided by stereo displacement and the static accommodation cue corresponding to the display depth commonly causes discomfort, known as vergence-accommodation conflict \cite{julesz1971foundations}.
The duration of pure vergence movements is influenced by travel distance, direction, and starting depth \cite{templin2014modeling}. 
Measurement of vergence movements are also more challenging compared to saccades due to the relatively smaller amplitude of movements \cite{yang2002latency,yang2004saccade}, inconsistent performance \cite{welchman2008bayesian}, complex neural coding \cite{cullen2011neural,semmlow2019dynamics,king2011binocular}, and a higher sensitivity to external factors such as pupil dilation \cite{nystrom2016pupil,jaschinski2016pupil,feil2017interaction}.

In the real 3D world, saccade and vergence movements are more commonly combined than isolated because of the 3D distribution of visual targets \cite{lang2014saccade,kothari2020gaze}.
Prior literature has demonstrated that, relative to pure vergence, these combined eye movements are accelerated by the addition of saccades  \cite{yang2004saccade,coubard2013saccade,erkelens1989ocular,collewijn1995voluntary,pallus2018response}. 
Competing theories attempt to untangle the neurological pathways that control vergence and combined movements, and fully explain their behaviors \cite{mays1984neural,zee1992saccade,quinet2020neural}.
However, there is no definitive and agreed-upon theory within the literature \cite{king2011binocular,cullen2011neural}, as exists for saccadic movements \cite{bahill1975main}.
Therefore, despite the critical importance of combined eye movements, we still lack an analytical understanding of how different vergence-saccade combinations quantitatively influence visual performance.
For instance, although adding a small saccade offset to a 3D target location may accelerate a slower vergence movement, would an extra long saccade provide even more acceleration, or would the benefits of the saccade be outweighed by additional travel time?
If so, what size saccade is optimal for producing the fastest vergence movement?
Our work attempts to answer these questions by quantifying the scale of this acceleration effect across different amplitudes of 3D gaze movements into a continuous domain probabilistic model for predicting gaze offset times, and side-step the need to explicitly depict the vast complexity of vergence-saccade movement behaviors.

\subsection{Stereo Vision and Stereopsis-Aware Optimization}
Understanding stereo vision in order to optimize computer graphics systems and user experience, especially in VR/AR environments, remains a popular research frontier \cite{aizenman2022statistics,shi2022stereo}.
Most of today's consumer VR/AR devices are incapable of supporting accommodation;
therefore, stereopsis is still the primary means by which these devices \emph{improve} depth perception over conventional 2D displays.%

Numerous efforts have been made to optimize stereoscopic content with gaze tracking so as to enhance the perceived realism of depth in virtual environments. Examples include grain positioning \cite{templin2014perceptually}, as well as optimizations considering depth \cite{kellnhofer2016gazestereo3d,templin2014modeling}, luminance \cite{wolski2022dark}, shading material \cite{chapiro2015stereo}, and displays \cite{zhong2021reproducing,chapiro2014optimizing}.
With the surge of low-cost and low-power gaze-tracking, another emerging research line incorporates dynamic cues such as motion parallax \cite{kellnhofer2016motion}. Depth cues may be enhanced by incorporating these various rotation and projection centers \cite{konrad2020gaze,krajancich2020optimizing}.
Reduced depth acuity in peripheral vision has also been leveraged to accelerate neural rendering \cite{deng2022fov} and image reconstruction \cite{kaplanyan2019deepfovea}.

\begin{figure*}[t]
    \centering
    \subfloat[setting and stimuli]{
        \includegraphics[height=3.89cm]{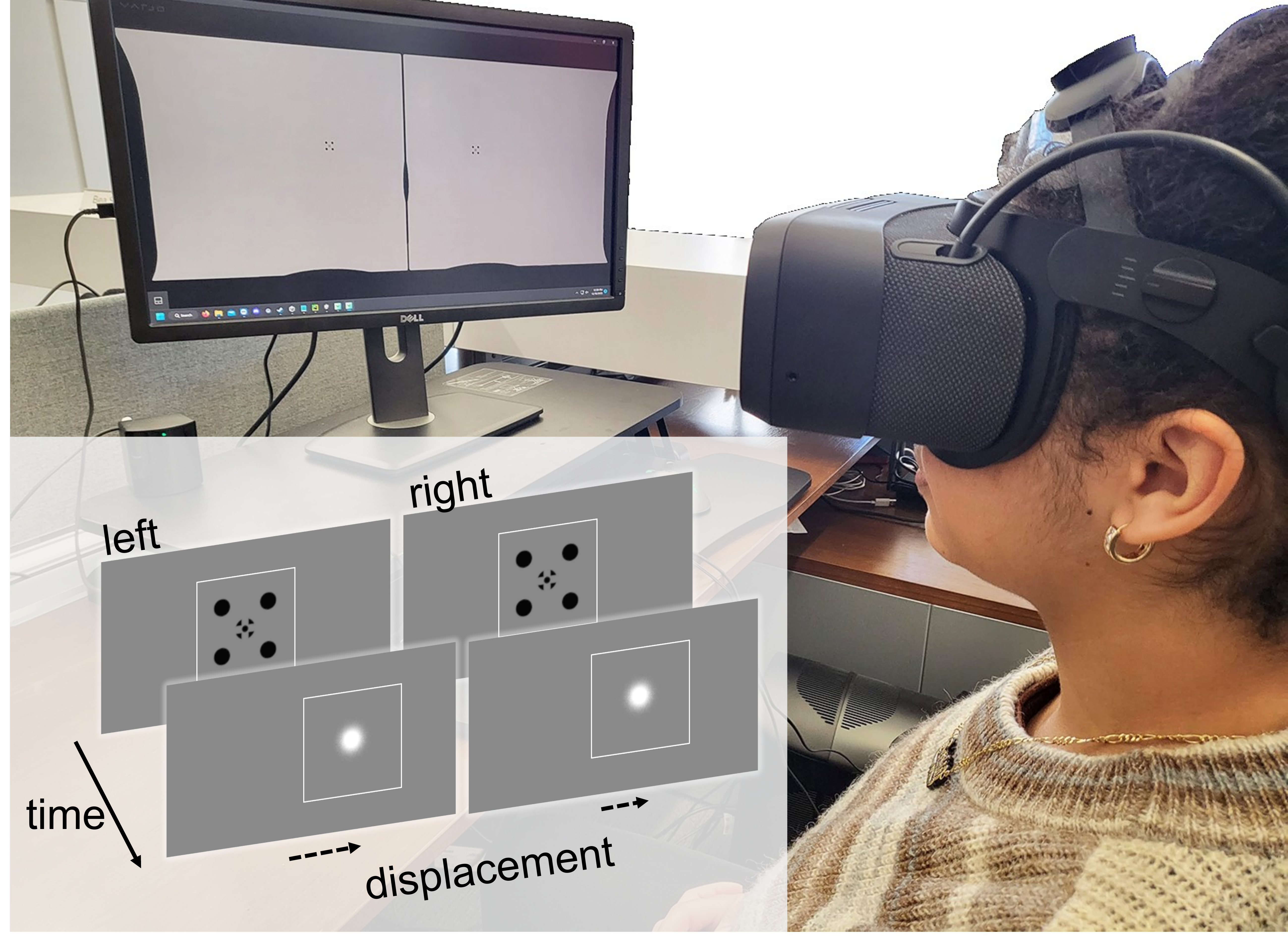}
        \label{fig:pilot:setting}
        \label{fig:pilot:stimuli}
    }%subfloat
    \subfloat[divergent]{
        \includegraphics[height=3.89cm]{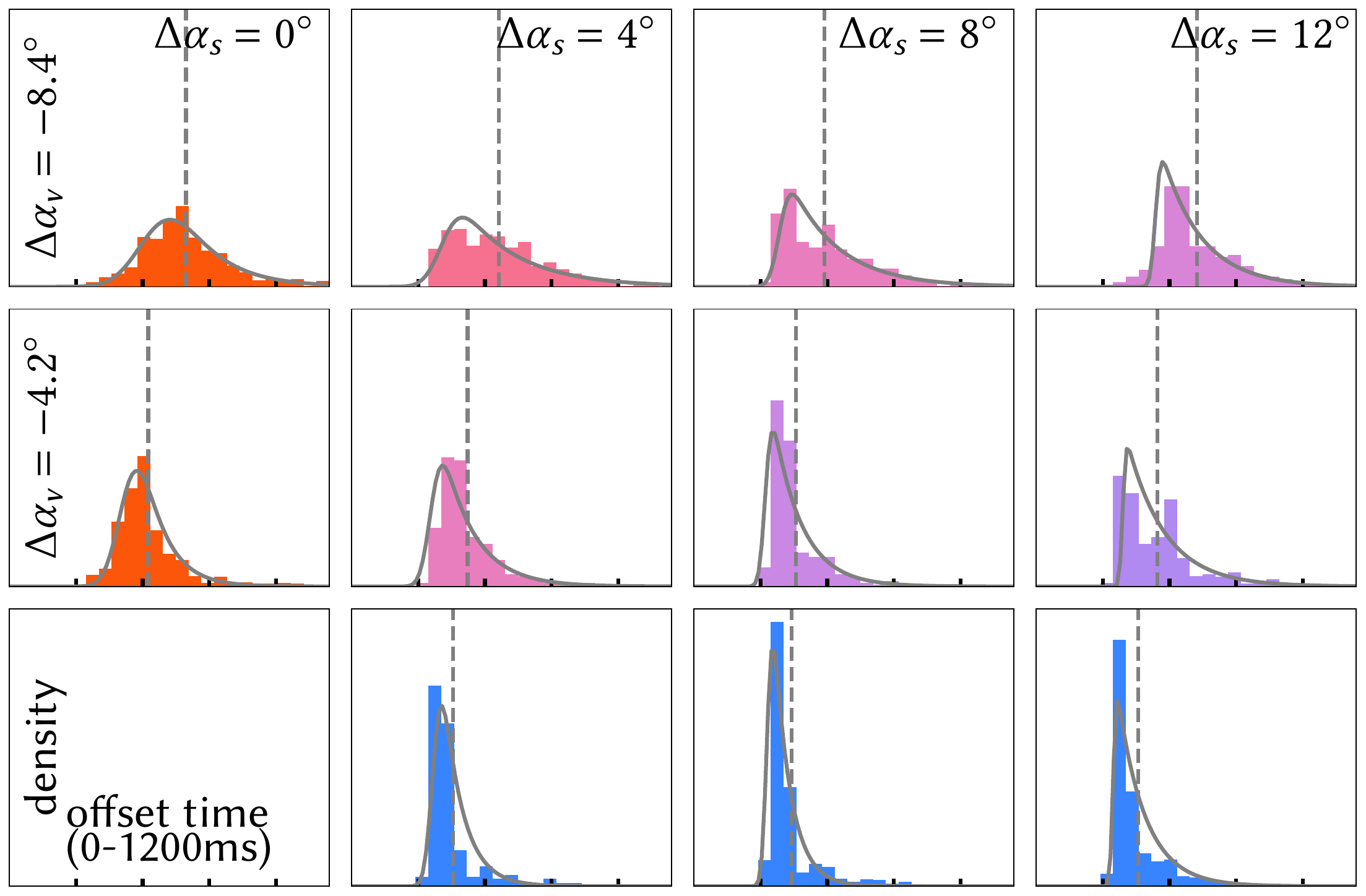}
        \label{fig:pilot:results:near}
    }%subfloat
    \subfloat[convergent]{
        \includegraphics[height=3.89cm]{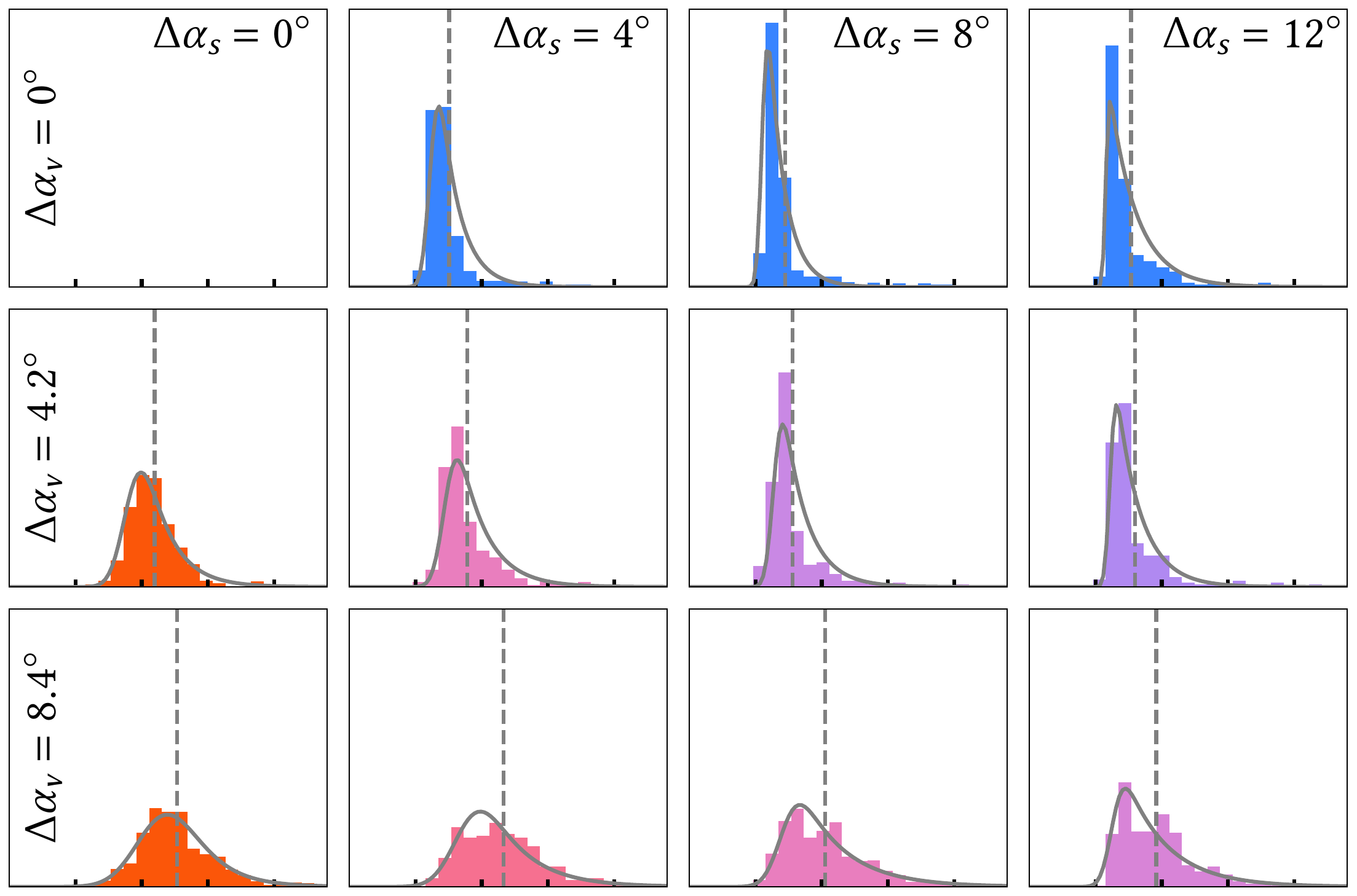}
        \label{fig:pilot:results:far}
    }%subfloat
    \Caption{
       Study setup and results.
    }{
    \subref{fig:pilot:setting} visualizes the setup and temporal stimuli (zoomed-in for illustration) of an example condition.
    \subref{fig:pilot:results:near}/\subref{fig:pilot:results:far} shows the histogram of the collected offset times, with divergent/convergent movement. 
    Each sub-figure block indicates an individual condition. Higher  vertical/horizontal locations imply higher vergence ($\vergentAmp$)/saccade($\saccadeAmp$) amplitudes.
    In each block, the X-axis denotes the observed offset time ($0-1200\ \text{ms}$ range; $250\ \text{ms}$ for each tick) and Y-axis denotes the corresponding distribution density.
    The dashed lines indicate the mean offset time of each histogram.
    For each histogram an Exponentially modified Gaussian (\emph{ExGauss}) distribution is fitted via Maximum Likelihood Estimation (MLE); refer to \Cref{sec:method} for details on the fitting procedure.
    }
\label{fig:pilot:results}
\label{fig:results:accuracy}
\end{figure*}

\section{Measuring and Predicting Stereoscopic Eye Movement Completion Time}
\label{sec:pilot}

To quantitatively understand combined stereoscopic eye movements, we first performed a psychophysical experiment with a wide field-of-view stereo VR display.
The study measured how jointly varying vergence and saccade amplitudes influence the time required for an observer's eyes to reach a 3D target relative to stimulus onset;
this duration is often referred to as the eye movement {\emph{offset time}}.
The data then serve as the foundation of our model (detailed in \Cref{sec:method}) for predicting the offset timing of various eye movements.

\subsection{Experimental Design}
\label{sec:pilot:design}

\paragraph{Participants and setup}

Eight participants (ages $20$-$32$, $6$ male) with normal or corrected-to-normal vision were recruited. 
Due to the demanding requirements, established low-level psychophysical research commonly starts with pilot studies involving a small number of participants and leverages the collected data to develop computational models (e.g., the foveated rendering literature \cite{patney2016towards,sun2020eccentricity,krajancich2021perceptual,krajancich2023towards}). 
These models, constructed using data from a limited set of subjects, can be evaluated for their cross-subject generalizability using a larger group of users, as we performed in \Cref{sec:results:study} with 12 additional unseen participants.
Moreover, in the context of our work, psychophysical studies examining the temporal dynamics of human behaviors require remarkably large sample sizes for a comprehensive statistical pattern to account for neural and mechanical noise \cite{yang2004saccade,collewijn1995voluntary,erkelens1989ocular,bucci2006latency,van2007sources}.
Considering that variations among subjects do not exhibit a significant impact on the completion rate of low-level gaze movements like saccades \cite{bahill1975main} and vergence movements \cite{collewijn1995voluntary,erkelens1989ocular}  – as confirmed by our cross-validation analysis in \Cref{sec:results:generalization} – and given that these are objective psychophysical behaviors not reliant on subjective reporting, we chose to enlist a small number of participants while acquiring an extensive sample size (1,500+ trials) per participant.
To this aim, we split the study across multiple days for every participant (see \emph{Conditions} paragraph for details).

The study was conducted with a Varjo Aero head-mounted VR display (HMD) with the relevant specifications detailed in \Cref{tab:pilot:hmd}.
As shown in \Cref{fig:pilot:setting}, throughout the study, participants wearing the HMD remained seated and performed the visual-target-changing task as detailed in the \emph{Task and Stimuli} paragraph.
Before the experiment, participants underwent a ``preamble'' checklist to ensure proper task completion and accuracy, including:
\begin{enumerate}[leftmargin=*]
    \item Measure and calibrate the HMD's inter-pupillary distance (IPD).
    \item Complete a five-point calibration for accurate binocular gaze tracking (repeat whenever the HMD is re-mounted after breaks).
    \item Adjust a fixation point between the nearest and furthest depths at which experimental stimuli appeared to ensure the success of fusing the stereoscopic visual stimuli (i.e., no double-vision).
\end{enumerate}
\paragraph{Task and stimuli}
Participants' task was to shift their gaze to land on targets appearing in 3D space. 
At the beginning of each trial, they were instructed to observe the fixation stimulus at the center of the screen.
As illustrated in \Cref{fig:pilot:stimuli}, this stimulus included a combination of a cross and four circular flankers to assist fixation \cite{thaler2013best}.
Once successful fixation was detected, this stimulus disappeared and was immediately replaced by a target stimulus, to which participants were instructed to move their gaze to as naturally as possible with a single gaze motion.
The target stimulus was a Gaussian blob with $\sigma = 0.25^{\circ}$ and peak luminance of $150\ \text{cd}/\text{m}^2$ --- a similar design as in Lisi~et~al.~\shortcite{lisi2019gain}.

To ensure stable tracking, a trial only began if the participant's eyes were within $1.2^\circ$ to the center of the fixation point for a consecutive $0.4\ \text s$.
If the participant failed to hold their gaze at the fixation point for sufficient duration more than three consecutive times, the eye-tracker was re-calibrated.
Additionally, to ensure correct task completion, we rejected and repeated a trial if it was completed in less than $0.1\ \text s$ or more than $1.3\ \text s$.
To avoid fatigue, participants were shown a darkened screen between trials as a cue to blink or close their eyes, if they: (1) successfully completed a trial, (2) failed to hold their gaze on the starting fixation point, or (3) failed a trial. 

\paragraph{Definitions and annotations}
Offset times are known to vary depending on the spatial location of the stimuli, mostly due to the varying contributions of either saccadic or vergence movements, often superimposed on each other \cite{zee1992saccade}.
In order to study how the spatial placement of the stimuli influences what type of eye movements arise, we parameterize spatial locations using two parameters: the vergence angle, $\vergentAng$, and the saccade angle, $\saccadeAng$, as illustrated in \Cref{fig:background:angles}.
All locations in the transverse plane containing the participants' eyes, and the stimuli can be encoded using the two degrees of freedom provided by $\vergentAng$ and $\saccadeAng$.

Specifically, following vision science practice, we define the vergence angle as the angle formed by the intersection of the gaze rays.
That is, if we denote the signed angles of the left and right eyes, with respect to the forward ``$z$'' direction (i.e. the intersection between the transverse and median planes) as $\leftAng$ and $\rightAng$, the vergence angle is equal to
\begin{align}
\vergentAng = {\leftAng - \rightAng}.
\end{align}
The set of gaze locations that have the same $\vergentAng$ form an \emph{isovergence circle}, visualized as the orange circles in \Cref{fig:background:angles}.
Pure vergence movements maintain the direction of gaze and move the gaze point from one isovergence circle to another.

On the other hand, the saccade angle, $\saccadeAng$, is defined as the mean of the angles of the left and right eyes:
\begin{align}
\saccadeAng = {(\leftAng + \rightAng)}/{2}.
\end{align}
The set of gaze locations that have the same $\saccadeAng$ form a ray representing the direction of gaze, visualized as the blue lines in \Cref{fig:background:angles}.
Pure saccadic movements remain on the same isovergence circle while rotating the direction of gaze across the transverse plane.

Therefore, a vergence and saccade angle pair, $\combineAng = (\vergentAng, \saccadeAng)$, unique\-ly defines a point on the transverse plane via the intersection of the isovergence circle which corresponds to $\vergentAng$, and the direction of gaze which corresponds to $\saccadeAng$.
An arbitrary gaze movement in this coordinate system can be represented as a displacement vector,
\begin{align}
\combineAmp
= \combineAng^\textit{t} - \combineAng^\textit{o}
= (\vergentAng^\textit{t} - \vergentAng^\textit{o}, \saccadeAng^\textit{t} - \saccadeAng^\textit{o})
= (\vergentAmp, \saccadeAmp),
\end{align}
for movement from $\combineAng^\textit{o(rigin)} = (\vergentAng^\textit{o}, \saccadeAng^\textit{o})$ to $\combineAng^\textit{t(arget)} = (\vergentAng^\textit{t}, \saccadeAng^\textit{t})$.

\paragraph{Conditions}
We define a condition by a pair $\{\combineAng^\textit{o}, \combineAmp\}$.
We sought to create a grid of experimental conditions which cover a wide set of possible gaze movements.
Today's VR devices limit the breadth of applicable eye movements.
Here we discuss these limitations as well as the solutions we implemented to ensure study accuracy.

First, we observed that participants could not fuse a stereo stimulus when it was placed too close, causing double (yet in-focus) vision.
This restricted the range of possible vergence movements we could study in VR.
We believe this effect is due to the lack of support for variable accommodation in VR displays, and thus distorted depth cues due to the \emph{vergence-accomodation conflict} \cite{hoffman2008vergence, aizenman2022statistics, march2022impact}.
To establish a conservative \emph{minimum} depth with successful stereo stimulus fusion, we performed a pre-study test with $4$ participants with various inter pupil distances (IPDs) ($64 - 71\ \text{mm}$).
Through this experiment, we established that this depth is approximately $\minDepth = 0.4\ \text m$ in front of the observer.
This corresponds to a \emph{maximum} vergence angle coordinate of $\vergentAng^\textit{max} = 8.4^\circ$ for an observer with an IPD of $\ipd^\textit{min} = 59\ \text{mm}$ --- the lowest IPD supported by the HMD (see \Cref{tab:pilot:hmd}).
Since a larger IPD only relaxes this maximum value, we limit the maximum vergence angle to $\vergentAng^\textit{max} \leq 8.4^\circ$.
See \Cref{sec:supp:pilot:conditions} for a more in-depth analysis.

Second, we found that the accuracy of the HMD eye tracker deteriorates significantly further in the periphery for $\saccadeAng \geq 15^\circ$.
We recognize that the majority of saccades naturally performed by humans have amplitudes $\saccadeAng \leq 15^\circ$ \cite{bahill1975most},
due to a preference to move the head otherwise.
Therefore, we limit the maximum saccade angle to $\saccadeAng^\textit{max} \leq 15^\circ$.

Lastly, due to the inconsistent nature of temporal human behavior, our study requires many repeats for each condition in order to reveal statistical trends. 
It is therefore infeasible to include a large number of conditions in our study.
We address this by only sampling gaze movement displacements, $\combineAmp$.
That is, although the initial gaze position $\combineAng$ has been shown to be a relevant factor influencing offset time \cite{templin2014modeling}, we chose not to consider it in our analysis and modeling for the current study. We leave characterizing the effects of ``starting pose'' as future work.

To summarize, our study design is constrained to vergence angles $\vergentAng \leq 8.4^\circ$, saccade angles $\saccadeAng < 15^\circ$, as well as to only consider gaze movement displacements, $\combineAmp$, and to ignore initial gaze positions, $\combineAng^o$.
Within these constraints, we sample the following conditions for vergence, saccade, and combined motions respectively:
\begin{itemize}[leftmargin=*]
    \item $2$ vergence conditions with amplitudes $(|\vergentAmp| \in \{4.2^\circ,$ $8.4^\circ\})$ conducted for both divergent ($-$) and convergent ($+$) movements,
    \item $3$ saccade conditions with amplitudes $(\saccadeAmp \in \{4^\circ, 8^\circ, 12^\circ\})$ conducted at near and far depths,
    \item $2 \times 3$ combined movements for every combination of the above conditions for both convergent and divergent movements,
\end{itemize}
totaling in $(2 + 3 + 2 \times 3) \times 2 = 22$ conditions, as in \Cref{fig:pilot:results:near,fig:pilot:results:far}.
We treated leftward and rightward saccades as symmetric;
therefore, while we randomized stimulus location to appear on the left or right side, in data processing, we remove the distinction by taking the absolute value of the saccade amplitudes.
Implementation of the conditions is detailed in \Cref{sec:supp:pilot:conditions}.

To account for human sensory and behavioral noise \cite{van2007sources}, we repeated each condition $6$ times within one experimental block (totaling in $6 \times 22 = 132$ trials per block), and instructed participants to complete a total of $12$ blocks.
Each block took $10-15$ minutes to complete, with a $2-3$ minute break between blocks.
The experiment was split into sessions across $3$ days to avoid fatigue, with each session scheduled at approximately the same time for consistent performance.
Before each session, participants also performed a short warm-up session of $24$ trials to familiarize themselves with the task and target positions and eliminate potential variance in reaction time.
Overall, each experimental condition was repeated a total of $72$ times, and the entire experiment took about $3$ hours for each participant, including intermediate breaks.
Running the experiment across $8$ participants, we collected a total of $8 \times 72 \times 22 = 12,672$ trials.

\paragraph{Data analysis}
Each experimental trial yields a time-series of eye directions recorded during the trial, sampled at $200\ \text{Hz}$.
Similar to \cite{yang2002latency,templin2014modeling,yang2010central}, we performed post-hoc processing and analysis on the raw data to more precisely identify gaze movement offset times.
To address tracker noise from high sampling frequency \cite{van2007sources}, we first applied a $25$ Hz smoothing filter \cite{butterworth1930theory}, similar to \cite{yang2010central,templin2014modeling}. 

We compute the angular velocity over time across each trial from the smoothed eye direction data and apply a constant velocity threshold to detect offset timestamps of gaze movement.
Specifically, for a reliable offset time measurement, we require two conditions to be met:
(1) individual speeds of the left and right eyes to be below a threshold of $5^\circ$/sec,
as well as (2) each eye to be directed within $1^\circ$ relative to the target.
While some prior work suggests that vergence offset times can be detected by the angular velocity in the vergence dimension, i.e.,
$
\frac{d}{dt} \vergentAng = \frac{d}{dt} (\leftAng - \rightAng)
$
\cite{yang2004saccade},
we found that our strategy is more fitting in our use case due to the additional challenges in eye tracker precision, accuracy, and frequency posed by consumer VR devices.
For consistency and fairness across all conditions, we applied this detection approach for all the conditions, including vergence-only, saccade-only, and combined movement trails.
A small percentage of trials ($6.4\%$) were rejected from analysis and training due to the gaze offset position falling outside the allowable range.
Manual inspection of these trials indicates that the users' eye movements only satisfied the second condition (2) above, but not the first (1).
These cases could not be identified during experiment run-time due to the inability to reliably perform post-processing filters to the raw data on the fly.

\subsection{Results}
\label{sec:pilot:results}
\begin{figure}[t]
    \centering
    \subfloat[saccade]{
        \vphantom{\includegraphics[width=0.290\linewidth,valign=t]{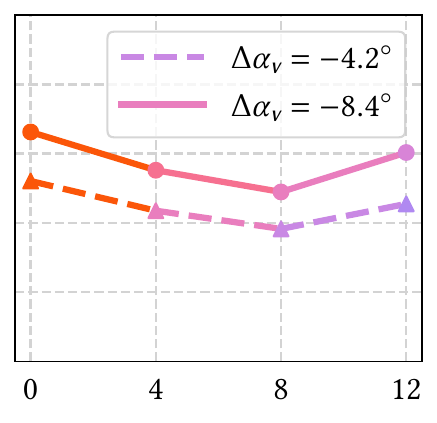}}
        \includegraphics[width=0.330\linewidth,valign=t]{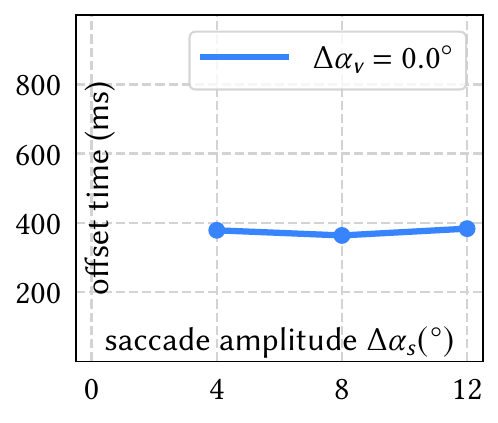} \label{fig:pilot:summary:saccade}
    }%subfloat
    \subfloat[divergent]{
        \includegraphics[width=0.290\linewidth,valign=t]{figures/pilot_data_aggregate_divergent.pdf}
        \label{fig:pilot:summary:divergent}
    }%subfloat
    \subfloat[convergent]{
        \includegraphics[width=0.290\linewidth,valign=t]{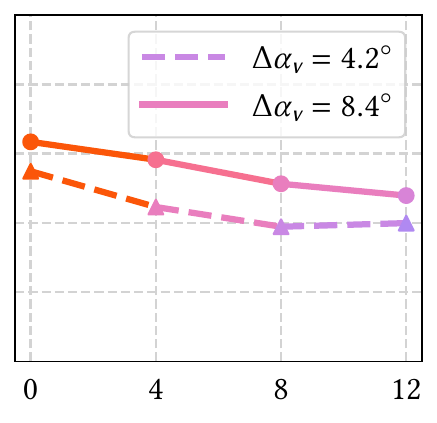}
        \label{fig:pilot:summary:convergent}
    }%subfloat
    \Caption{Aggregated mean offset time of studied conditions across all participants.
    }{%
\subref{fig:pilot:summary:saccade} shows the mean offset time of pure saccade conditions. 
X- and Y-axes indicate saccade amplitudes, $\saccadeAmp$, and mean offset time, respectively (offset time std shown in \Cref{sec:supp:results}). 
Note the consistency across varied amplitudes.
\subref{fig:pilot:summary:divergent}/\subref{fig:pilot:summary:convergent} show the mean offset times with pure vergence ($\saccadeAmp=0$) and combined movement ($\saccadeAmp\neq0$) conditions. 
Note the non-monotonic/u-shaped effect of $\saccadeAmp$ on the offset time.
    }
\label{fig:pilot:summary}
\end{figure}

\Cref{fig:pilot:results} visualizes the raw data with the identified eye movement offset time. All time values in the statistical analysis below and throughout the paper are in \emph{seconds} for clarity.
Additionally, \Cref{fig:pilot:summary} statistically summarizes the mean of each condition.

The offset times of saccades ($\vergentAmp=0^\circ$, $.37\ (\text{mean}) \pm .12\ (\text{std})$) are lower than offset times of vergence movements ($\saccadeAmp=0^\circ$, $.59 \pm .15$).  
The effect applies for both divergent ($\vergentAmp<0^\circ$, $.59 \pm .17$) and convergent ($\vergentAmp>0^\circ$, $.59 \pm .14$) conditions. 
The average offset time of combined movements ($.48\pm .16$) lies in between. 
A repeated measures analysis of variance (ANOVA) indicated that the type of eye movement (saccade/vergence/combined) had a significant effect on the offset time ($F_{2,14}=339.3, p<.001$). 
Additionally, the range (max-min) of mean offset times across saccade conditions ($.02$) is significantly narrower than across vergence conditions ($.14$). 
The effect can be visualized by comparing the span of values on the Y-axis of \Cref{fig:pilot:summary}.

Larger vergence amplitudes ($|\vergentAmp|$) significantly prolong the offset time in combined movements. 
For example, the average landing time for $|\vergentAmp|=4.2^\circ/8.4^\circ$ is $.53\pm.12$/$.65 \pm .16$. 
A repeated measures ANOVA indicated that the $|\vergentAmp|$ had a statistically significant effect on the offset time ($F_{2,14}=384.7, p<.001$). 

For combined offset times, we did not observe a monotonic effect of saccade amplitude ($\saccadeAmp$).
In fact, with a given vergence amplitude, the effect of saccade amplitude on the combined movement time is inconsistent and commonly non-monotonic, as visualized with the ``U-shape'' in \Cref{fig:pilot:summary:divergent}.
The average landing time for pure saccade conditions, $\saccadeAmp=4^\circ/8^\circ/12^\circ$, are $.38 \pm .12$/$.36 \pm .11$/$.38 \pm .13$.
When $\vergentAmp=-8.4^{\circ}$, however, the fastest combined movement occurs for $\saccadeAmp=8^{\circ}$ ($.49\pm .16$), compared with the other two conditions $\saccadeAmp=4^{\circ}$ ($.55 \pm .18$) and $\saccadeAmp=12^{\circ}$ ($.60 \pm .15$).
A Mann-Kendall trend test did not observe a significant monotonic trend ($\tau=.33, p=1.0$).

The distribution of offset times across all conditions exhibits positive skewness ($\gamma_1 = 1.94 \pm .89$).
Among the conditions, skewness varied by condition with pure vergence movements is the smallest ($1.4$), combined movements in the middle ($1.8$), and pure saccadic movements is the highest ($3.1$).
This indicates that different gaze movements change the shape of the distribution of offset times, which can also be visualized from the histograms in \Cref{fig:pilot:results}.

\subsection{Discussion}
\label{sec:pilot:discussion}
The visualization and analysis draw us to several conclusions.
First, the offset times of singular saccade movements are significantly shorter and more consistent than those of vergence movements.
Second, statistical analysis of our data evidenced that slow vergence movements are ``accelerated'' if combined with faster saccades. 
Third, the acceleration effect varies depending on how they are combined.
Saccade acceleration exhibits a ``U-shape'' for divergent combined movements (\Cref{fig:pilot:summary:divergent}).
The optimality (i.e., the amplitude of the saccade that accelerates vergence the most, thus the fastest combined movement) depends on the corresponding vergence amplitude.
Lastly, human performance on changing 3D visual targets is inconsistent across trials, even within the same participant.
Moreover, the scale of the inconsistency varies across different eye movements.
These observations inspire us to develop a computational model that 1) depicts quantitatively how saccades accelerate vergence, and 2) predicts the probability distribution of target landing offset time with combined vergence-saccade movements.

\subsection{Generalization to Arbitrary Gaze Movements}
\label{sec:method}

\paragraph{Statistical model}
The statistical analyses in \Cref{sec:pilot:results,sec:pilot:discussion} motivate us to develop a model for predicting the target landing offset times for arbitrary gaze movements not present within our dataset.
As reported in \Cref{sec:pilot:results}, the distributions observed in our dataset are positively skewed, and vary across different conditions;
so an Exponentially modified Gaussian (\emph{ExGauss}), which features fine control over skewness via its parameters, is a viable choice of statistical model for these distributions \cite{marmolejo2023generalised}.
Specifically, offset time, $\travelTime$, represented as an \emph{ExGauss} random variable has a probability density function (PDF),
\begin{align}
\begin{aligned}
\PDF_\travelTime(\timestamp; \mean, \std^2, \decay)
&=
    \frac1{2\decay}
    e^{2\mean + \frac{\sigma^2}\decay - 2\timestamp}
    \erfc\left(
         \frac{\mean + \frac{\sigma^2}\decay - \timestamp}{\sqrt2\sigma}
     \right),
\end{aligned}
\label{eqn:exgausspdf}
\end{align}
parameterized by $\mean$, $\std$, and $\decay$, to depict the location, spread, and asymmetry of the resulting distribution, respectively.
All parameters are in units of \emph{seconds}.
Here, $\erfc(\cdot)$ is the complementary error function.
As shown in \Cref{fig:pilot:results}, we estimate the \emph{ExGauss} parameters for each condition separately via Maximum Likelihood Estimation (MLE) to collect a total of $N = 19$ sets of parameters (not double counting the saccade conditions).

In this work, offset times are modeled as \emph{ExGauss} random variables, but note that modeling with a different random variable may also be valid.
We leave the analysis and comparisons among model choices as future work since the specific presentation is beyond our focus, and other parameterizations are adaptable to our framework.

\paragraph{Parameter interpolation}
Our focus, instead, is on how the parameters of a given model should be interpolated to provide predictions of gaze offset times for arbitrary gaze movements.
To this end, we leverage the \emph{ExGauss} parameter estimations of each condition and smoothly interpolate each parameter via Radial Basis Function (RBF) interpolation.
Concretely, each RBF takes, as input, the amplitude of the gaze movement, $\combineAmp = (\vergentAmp, \saccadeAmp)$, to output the predicted \emph{ExGauss} random variable, $\travelTime(\combineAmp)$, with estimated parameters
\begin{align}
\begin{aligned}
\hat\mean(\combineAmp) &\coloneqq \sum_i^M \rbfWeight^\mean_i \varphi(\varepsilon^\mean||\combineAmp - \rbfCenter^\mean_i||),\\
\hat\std(\combineAmp) &\coloneqq \sum_i^M \rbfWeight^\std_i \varphi(\varepsilon^\std||\combineAmp - \rbfCenter^\std_i||),\\
\hat\decay(\combineAmp) &\coloneqq \sum_i^M \rbfWeight^\decay_i \varphi(\varepsilon^\decay||\combineAmp - \rbfCenter^\decay_i||).
\label{eqn:method:rbfMean}
\end{aligned}
\end{align}
$\rbfCenter^\mean_i$ and $\rbfWeight^\mean_i$ represent the location and weight of each of the $M = 4$ radial bases, $\varphi$ is the radial function, and $\varepsilon^\mean$ is a tuning shape parameter for the radial function.
In our implementation, we used the Gaussian kernel, $\varphi(r) = \exp(-r^2)$.
Overall, the learnable parameters in this regression are $\rbfCenter^j_i$, $\rbfWeight^j_i$, and $\varepsilon^j$ for $i \in \left[1 \dots M \right]$, totalling in $4 + 4 + 1 = 9$ variables for each \emph{ExGauss} parameter $j \in \left\{ \mean, \std, \decay \right\}$.

\paragraph{Regression}
We optimize the adjustable variables via gradient descent to minimize the mean-squared error between the MLE-estimated \emph{ExGauss} parameters for each condition, and the RBF-interpolated parameters, with the loss 
\begin{align}
\begin{aligned}
\loss_j &= \frac1N \sum^N \left( j - \hat j \right)^2 \text{ for } j \in \{ \mean, \std, \decay \}.
\label{eqn:method:loss}
\end{aligned}
\end{align}
The RBF parameters are regressed using batch gradient descent with the loss functions from \Cref{eqn:method:loss} and a learning rate of $10^{-2}$ for $200,000$ iterations.
The mean-squared losses are minimized from $137k/2.3k/17k\ \text{s}^2$ to $230/200/120\ \text{s}^2$ over the course of each regression, respectively.
We report model performance metrics as well as additional evaluations in \Cref{sec:result}.

\begin{figure}
  \centering
  \includegraphics[width=0.75\linewidth]{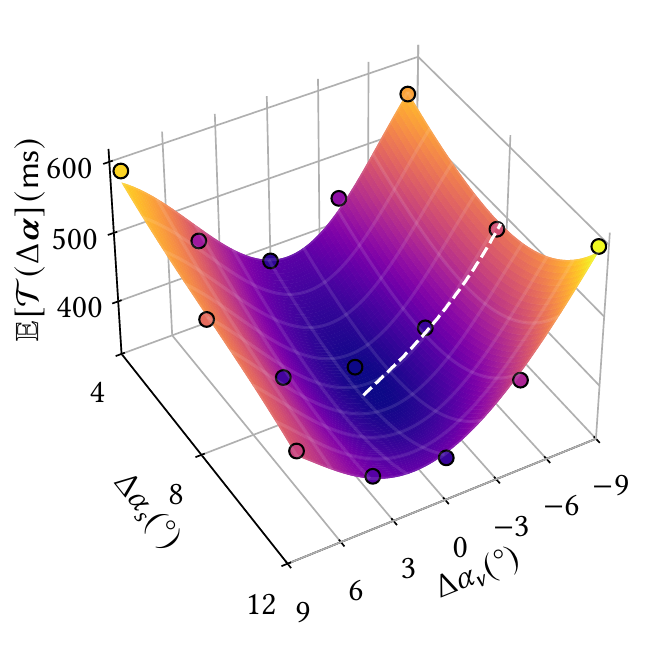}
  \Caption{
      Visualization of the interpolated model.
  }{%
      The sparsely sampled data visualized in \Cref{fig:pilot:summary} is smoothly interpolated via RBF interpolation.
      The surface heatmap shows the mean offset times across all interpolated conditions, and the measured data is overlaid as a scatter plot for comparison.
      The ``optimal'' combined gaze movements at various vergence amplitude settings are computed using \Cref{eqn:method:optmize} and visualized as a dashed white line on the surface of the model prediction.
  }
  \label{fig:model}
\end{figure}

\paragraph{Discussion and applications}
We compare the mean offset times predicted by our model to the means aggregated from our dataset in \Cref{fig:model}.
This visualization demonstrates how the offset times differ between convergent and divergent gaze movements.
For convergent combined movement, we observe the same monotonic decrease in offset time as a function of saccade amplitude as reported in \Cref{fig:pilot:summary:convergent}.
Additionally, we see the U-shaped behavior for divergent combined movements, as discussed in \Cref{sec:pilot:discussion,fig:pilot:summary:divergent}.

The \emph{ExGauss} distribution and RBF interpolation methods are represented by parameterized differentiable functions.
This allows us to compose these components to construct an end-to-end differentiable model for predicting the probability distribution of arbitrary gaze movements.
This formulation can be leveraged in various ways for practical applications.
For example, the ``optimal'' saccade amplitude, $\saccadeAmp^*$, which minimizes the offset time at various vergence amplitudes, $\vergentAmp$ can be computed analytically:
\begin{align}
\begin{aligned}
\saccadeAmp^*
&= \argmin_{\saccadeAmp} \expectation\left[\travelTime\left(\combineAmp = \left(\vergentAmp, \saccadeAmp\right)\right)\right]\\
&= \argmin_{\saccadeAmp}\left(\hat\mean\left(\vergentAmp, \saccadeAmp\right) + \hat\decay\left(\vergentAmp, \saccadeAmp\right)\right).
\label{eqn:method:optmize}
\end{aligned}
\end{align}
These local minima indicate the location of the lowest point in the valley of the U-shaped behavior, as visualized in \Cref{fig:model}.

\section{Evaluation}
\label{sec:result}
We first measure the statistical accuracy and necessity of the ver\-gence-saccade combined modeling with an ablation study in \Cref{sec:results:accuracy}.
We further test the model's goodness-of-fit when generalizing to unseen users and trials in \Cref{sec:results:generalization}.
Then, to evaluate its applicability in real-world scenarios and novel conditions, we perform an evaluation user study with various scenes in \Cref{sec:results:study}.

\subsection{Model Accuracy and Ablation Study}
\label{sec:results:accuracy}
\label{sec:results:ablation}

\paragraph{Metrics}
We utilize the Kullback–Leibler divergence (\KLdiv) as a continuous domain metric for measuring the similarity between model-predicted probability densities and the histograms obtained from the psychophysical data.
A model with \emph{lower} \KLdiv relative to a ground truth histogram indicates a \emph{better} prediction. 

\begin{table}[b]
  \caption{KL divergence of the model and ablation study.}
  \label{tab:eval:ablation}
  \label{tab:eval:accuracy}
  \begin{tabular}{c|ccc}
    \toprule
    Condition & $\conditionAccuracyFull$ & $\conditionAccuracyVergence$ & $\conditionAccuracySaccade$\\
    \midrule
    KL Divergence & $.172$ & $.236$ & $.444$\\
    \bottomrule
  \end{tabular}
\end{table}

\paragraph{Conditions}
We conduct an ablation study and utilize the \KLdiv to validate the necessity of modeling combined movements.
Specifically, we consider the model's prediction accuracy if not supplying it with information on either saccade or vergence movement.
For this purpose, we re-aggregate our psychophysical data into groups separated only by saccade amplitude ($\conditionAccuracySaccade$), or only by vergence amplitude ($\conditionAccuracyVergence$) conditions.
That is, we pool together the histograms in \Cref{fig:pilot:results,fig:pilot:results} across the columns, or rows respectively.
The re-aggregation is then utilized to regenerate an ablated model following the same steps as described in \Cref{sec:method}.
See \Cref{sec:supp:ablation} for visualizations of the ablated model predictions.

While the probability distribution predicted by our model is continuous, the psychophysical study dataset only provides a finite sample of the theoretical ground truth distribution of offset times.
Therefore, we apply the discrete version of \KLdiv onto histograms of the ground truth data for each condition with $n = 50$ bins ($\Delta t = 24\ \text{ms}$).

\paragraph{Results and discussion.}
The resulting average \KLdiv{s} for the two ablated models are compared to the full model ($\conditionAccuracyFull$) in \Cref{tab:eval:accuracy}.
We observe that the $\conditionAccuracyFull$ model exhibits significantly lower \KLdiv than $\conditionAccuracyVergence$ and $\conditionAccuracySaccade$.
While the number of bins does have an effect on the divergence values, we extensively tested and confirmed that the relative relationship across the three conditions was not influenced by this factor.
These results demonstrate that combined eye movements exhibit remarkably distinct temporal patterns that depend both on saccade and vergence movement amplitudes, agreeing with our observations in \Cref{sec:pilot:discussion}.
Quantitatively, the combined model predicts participants' behaviors significantly more accurately, and thus proves the necessity and effectiveness of considering amplitudes of both components of movement.

\subsection{Model Generalizability}
\label{sec:results:generalization}
We further evaluate generalized goodness-of-fit with unseen data partitions.
We create segments of the psychophysical data from \Cref{sec:pilot} into training-testing groups along multiple axes.

\paragraph{Metrics}
Similar to prior art on stochastic visual behaviors \cite{Duinkharjav:2022:IFI,le2017visual}, we utilize the Kolmogorov-Smirnov (K.S.) goodness-of-fit test \cite{massey1951kolmogorov} between the test set and the corresponding model prediction, using ten quantiles for the offset time. 
Significance ($p<.05$) in the K.S. test indicates a rejection of the null hypothesis that two samples are drawn from the same distribution; failing to reject ($p>.05$) supports distributional matching.
The $D$ value in K.S. measures the maximum distance.

\paragraph{Conditions}
We first assess the model's statistical goodness of fit for the full set of psychophysical data from \Cref{sec:pilot}. 
Then we analyze the model's generalizability based on its capability to successfully fit the statistical distribution with unseen trials or subjects. 
To this end, the collected dataset is split into two fully separated training and testing sets without overlap.
The training set is leveraged to re-train a new model as in \Cref{sec:method}, which tests the fitness on the corresponding unseen test set.
We experiment with two methods of partitions:
(1) reserve each one of the eight participants' data as the test set (annotated as $\conditionEvalAccuracy_i,i\in\{1,2,\dots,8\}$;
(2) uniformly randomly sample 1/8 of the entire data for each condition but across all users (annotated as $\conditionEvalAccuracy_r$).
For both methods, the remaining data is used as the corresponding training set. 

\paragraph{Results and discussion}
\begin{figure}[t]
  \subfloat[K.S. statistics]{
    \definecolor{C1Color}{HTML}{1F77B4} 
\definecolor{C2Color}{HTML}{FF7F0E} 
\definecolor{C3Color}{HTML}{2CA02C} 
\definecolor{C4Color}{HTML}{D62728} 
\definecolor{C5Color}{HTML}{9467BD} 
\definecolor{C6Color}{HTML}{8C564B} 
\definecolor{C7Color}{HTML}{E377C2} 
\definecolor{C8Color}{HTML}{7F7F7F} 
\definecolor{CrColor}{HTML}{BCBD22}

\begin{tabular}{c|cc}
\toprule
& D & p \\
\midrule
\color{C1Color}{$\conditionEvalAccuracy_1$} & .1 &   1 \\
\color{C2Color}{$\conditionEvalAccuracy_2$} & .1 &   1 \\
\color{C3Color}{$\conditionEvalAccuracy_3$} & .2 & .99 \\
\color{C4Color}{$\conditionEvalAccuracy_4$} & .2 & .99 \\
\color{C5Color}{$\conditionEvalAccuracy_5$} & .1 &   1 \\
\color{C6Color}{$\conditionEvalAccuracy_6$} & .1 &   1 \\
\color{C7Color}{$\conditionEvalAccuracy_7$} & .1 &   1 \\
\color{C8Color}{$\conditionEvalAccuracy_8$} & .1 &   1 \\
\color{CrColor}{$\conditionEvalAccuracy_r$} & .1 &   1 \\
\midrule        
\end{tabular}

    \label{tab:eval:generalization:stats}
   }
   \hspace{0.3cm}
   \subfloat[Q-Q visualization]{
        \raisebox{-.5\height}{\includegraphics[width=0.53\linewidth]{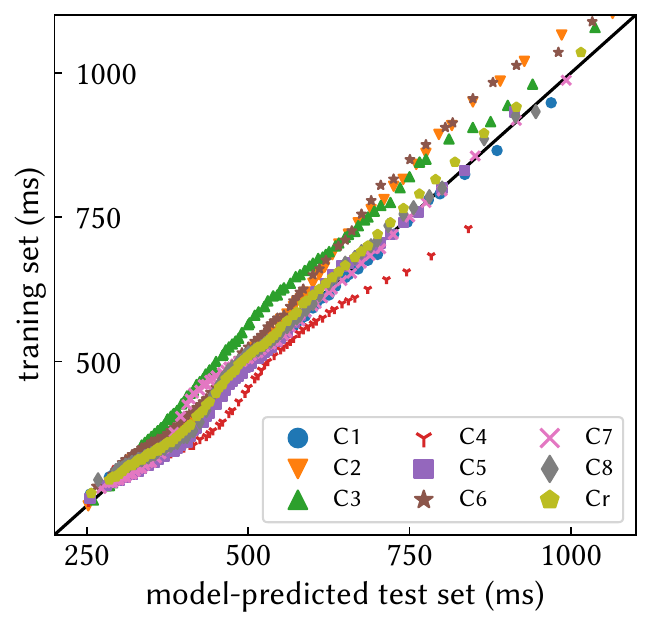}}
        \label{fig:results:generalization:qq:user}
    }%subfloat
    \Caption{Results of the model generalization evaluation with various partition conditions.}
    {%
    \subref{tab:eval:generalization:stats} shows the K.S. analysis. The color indicates the corresponding partition condition.
    \subref{fig:results:generalization:qq:user} shows the Q-Q plot for all conditions, comparing the distributions between the model-prediction on test set vs. training set.
    }
  \label{tab:eval:generalization}

\end{figure}

\Cref{tab:eval:generalization:stats} shows the results for the good\-ness-of-fit across all conditions. Additionally in \Cref{fig:results:generalization:qq:user}, we provide a quantile-quantile (Q-Q) visualization between the training set and the model prediction on the test set: samples closer to the diagonal line indicate better distribution agreement.
As a baseline reference, the K.S. test between the model and all collected data shows $D= .1, p=1$.
For all experimented partitioning conditions, the K.S. tests exhibit $p>.99$, 
failing to reject the null hypothesis that the model prediction acquired from the training set and the unseen test data are drawn from the same distribution.
The goodness-of-fit analyses above reveal that our probabilistic model can be generalized to unseen users and trials, implying that it can predict user behavior without observing it in advance.

\subsection{Study: Predicting and Optimizing Visual Performance}
\label{sec:results:study}

Beyond measuring the performance of the model on data from the controlled experiment (\Cref{sec:pilot}), we further design and conduct a second study with more complex stimuli.
We aim to gauge the model's capability to predict and optimize visual performance with realistic VR/AR scenarios, novel conditions, and unseen participants.

\paragraph{Participants and setup}
We recruited $12$ participants (ages $20-33$, $3$ female).
To validate the generalizability of the model, we ensured no overlap of participants with the study from \Cref{sec:pilot}.
All participants reported having normal or correct-to-normal vision.
We utilized the same hardware and ``preamble'' checklist as in \Cref{sec:pilot:design}.

\begin{figure}[tb]
    \centering
    \subfloat[archery \& 2D target]{
        \includegraphics[width=0.305\linewidth]{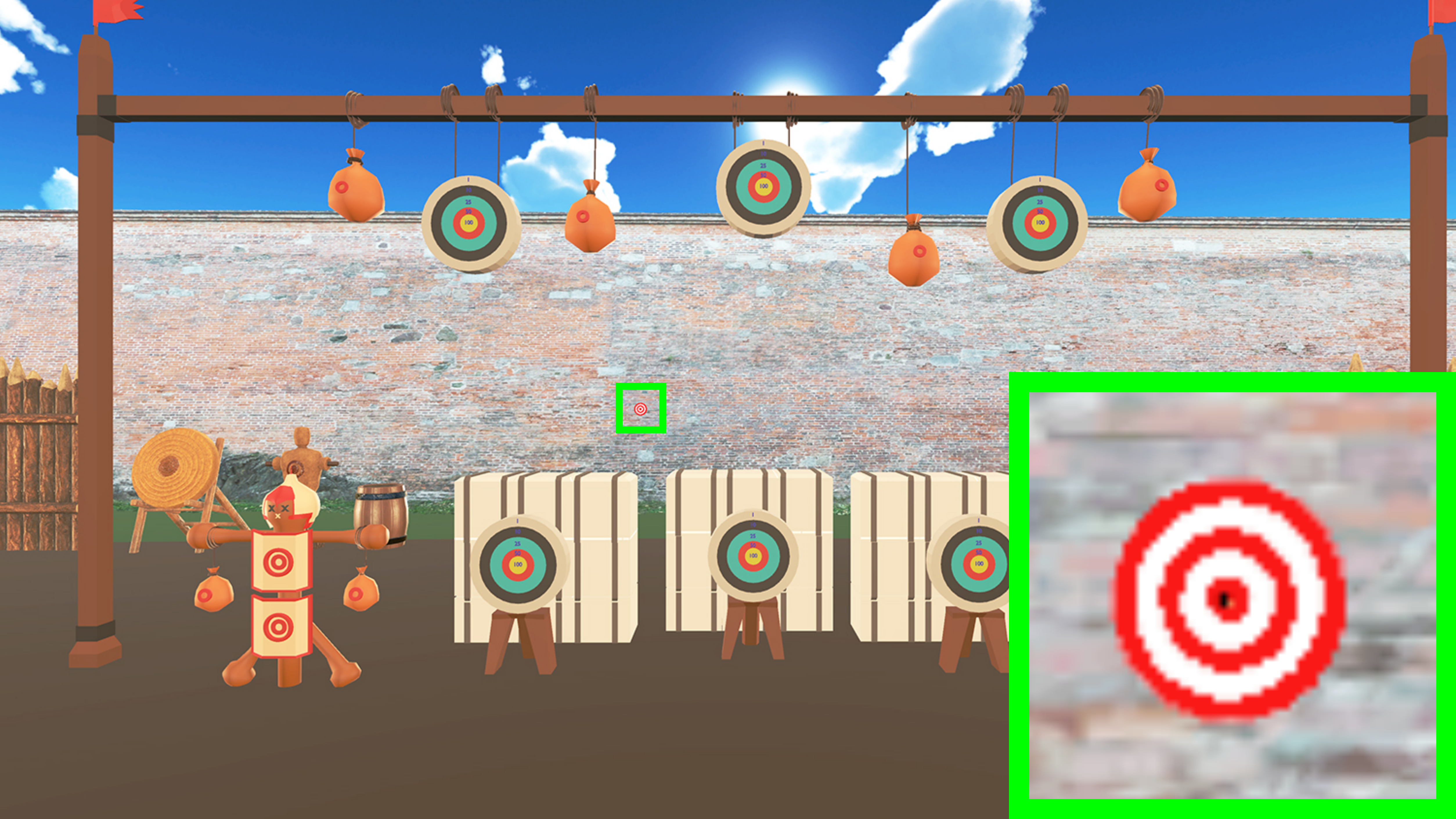}
        \label{fig:results:study:scene:archery}
    }%subfloat
    \subfloat[basketball \& 3D target]{
        \includegraphics[width=0.305\linewidth]{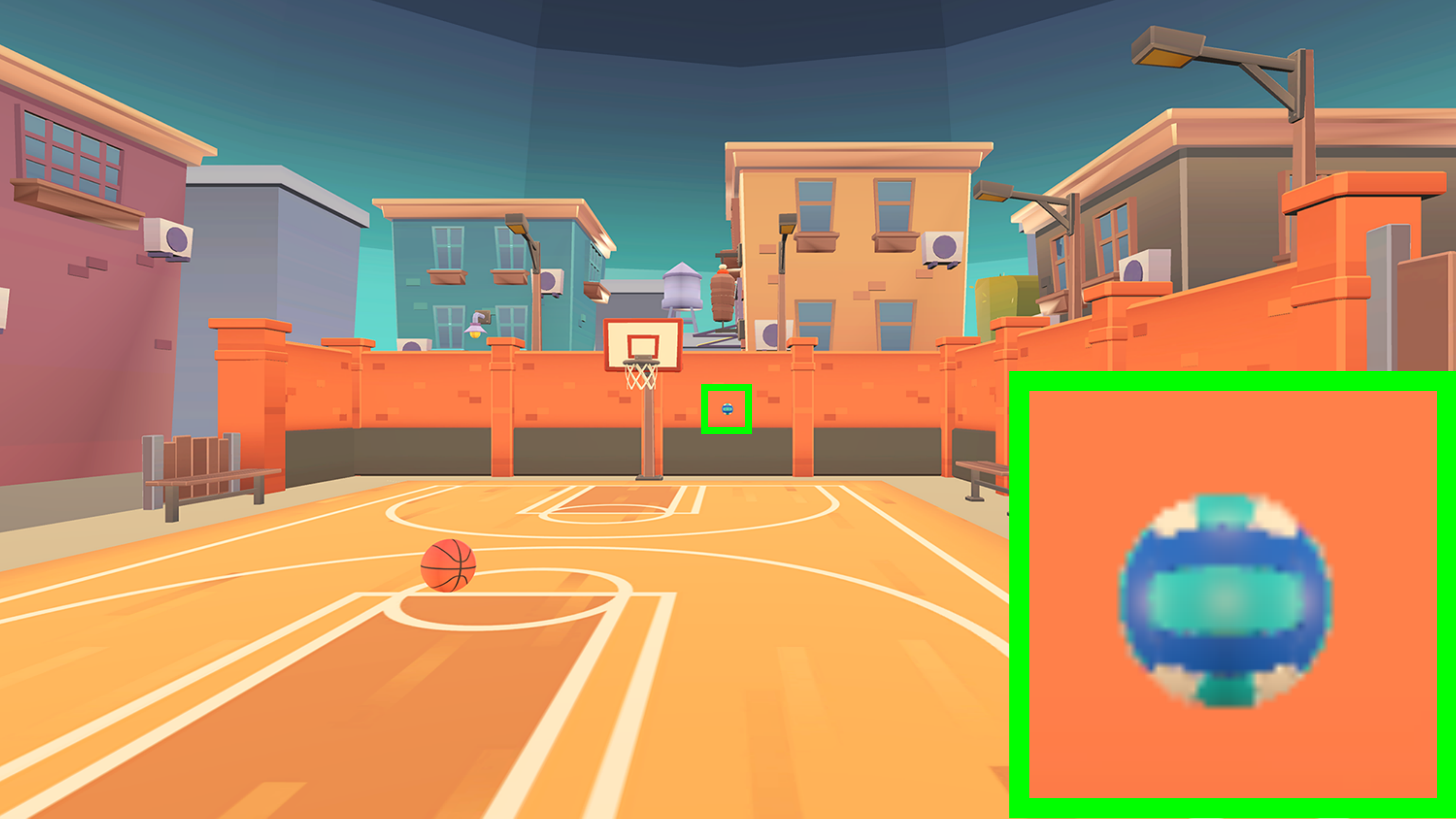}
        \label{fig:results:study:scene:bball}
    }%subfloat
    \subfloat[natural]{
        \includegraphics[width=0.305\linewidth]{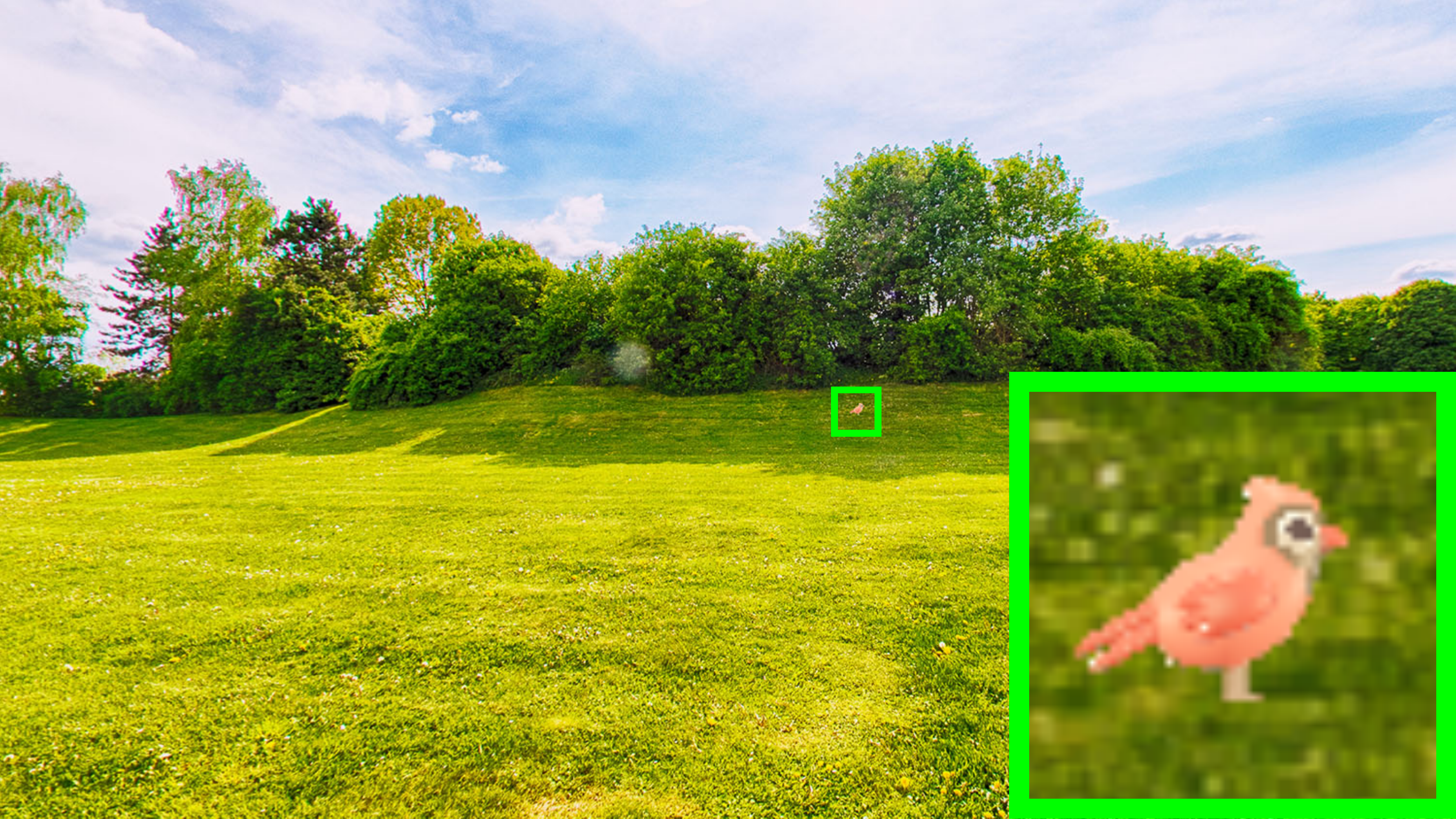}
        \label{fig:results:study:scene:ar}
    }%subfloat

    \subfloat[by conditions]{
        \includegraphics[height=3.7cm,valign=t]{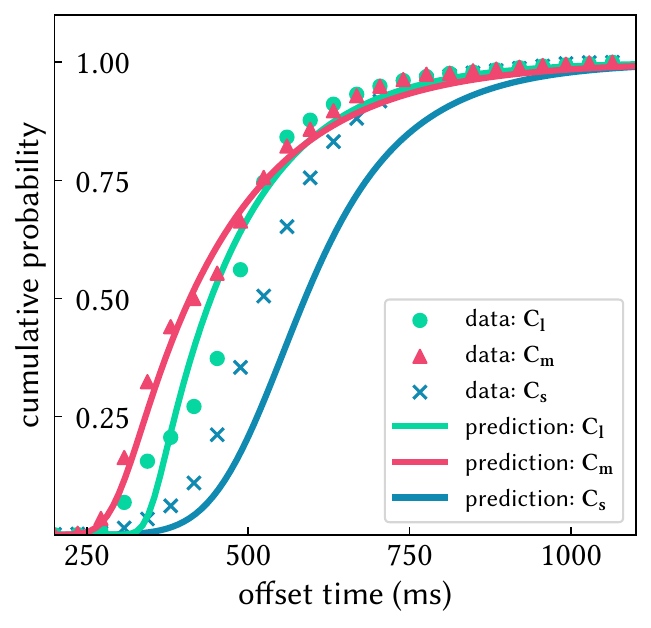}
        \label{fig:results:study:fit:all}
    }%subfloat   
    \subfloat[by participants]{
        \includegraphics[height=3.7cm,valign=t]{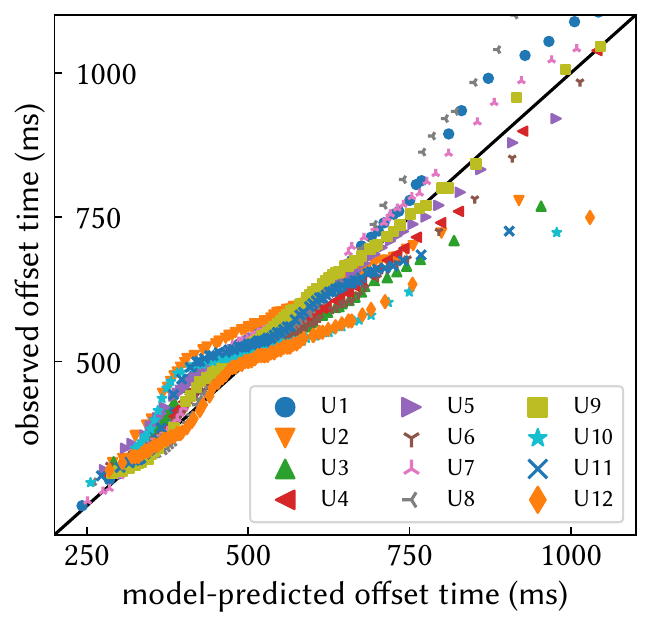}
        \label{fig:results:study:fit:user}
    }%subfloat   
    \Caption{
       Evaluation user study scenes and results.
       }{
       The first row shows the $3$ scenes leveraged for the study. The target stimuli are zoomed-in with insets. 
       The second row visualizes the comparisons across various dimensions.
       \subref{fig:results:study:fit:all} compares the model vs. data for the $3$ conditions, aggregating all users and scenes. The X-axis/Y-axis indicates offset time/cumulative probability. Note the  discrepancy between eye travel distance ($\conditionEvalStudyVergent<\conditionEvalStudyCombinedShort<\conditionEvalStudyCombinedLong$) and landing time ($\conditionEvalStudyCombinedShort<\conditionEvalStudyCombinedLong<\conditionEvalStudyVergent$).
       Predictions for $\conditionEvalStudyVergent$ appear higher than measured data, but are statistically similar (\Cref{sec:results:study}).
       \subref{fig:results:study:fit:user} visualizes the model vs. data for each of the participants with a Q-Q plot, aggregating all conditions and scenes. Samples closer to the diagonal line indicate better fitting.
       }
    
\label{fig:results:study}
\end{figure}

\paragraph{Scenes and stimuli}
To validate how our model performs for varied scenarios and content, we designed $3$ distinct environments: 
(1) a rendered archery range with a 2D bullseye stimulus (\Cref{fig:results:study:scene:archery}), 
(2) a rendered basketball court with a 3D ball stimulus (\Cref{fig:results:study:scene:bball}), and 
(3) a photographic natural outdoor scene with a virtual bird stimulus to simulate pass-through augmented reality (AR) scenarios (\Cref{fig:results:study:scene:ar}).

\paragraph{Tasks}
We instructed participants to complete a target-changing task similar to \Cref{sec:pilot:design}.
During each trial, participants were first instructed to fixate on a cross at the center of the screen.
After successfully fixating for $0.4\ \text s$, the cross was immediately replaced by one of the three scenes, containing the corresponding target at a new location. 
The participant then made an eye movement to direct their gaze at the target stimulus.
To reduce the influence of progressive learning effects on reaction time, as well as to familiarize the participants with the environment and task, participants performed $36$ warm-up trials for each of the scenes, followed by a short break.

\paragraph{Conditions}
We aim to validate our realistic scenarios with unseen conditions during the model training. 
Given the hardware limitations in \Cref{sec:pilot:design}, we experimented with a fixation at $0.4\ \text{m}$ and targets placed $\vergentAmp=6.9^\circ$ away in depth.
Using this novel vergence depth, we designed $3$ conditions with various eye travel distances:
\begin{enumerate}
    \item[$\conditionEvalStudyVergent$:] pure vergence motion with the \textbf{shortest} distance, $\saccadeAmp=0^\circ$, 
    \item[$\conditionEvalStudyCombinedShort$:] combined motion with the \textbf{medium} distance $\saccadeAmp=7^\circ$,
    \item [$\conditionEvalStudyCombinedLong$:] combined motion with the \textbf{longest} distance $\saccadeAmp=10.5^\circ$. 
\end{enumerate}
We used the same conditions across all three tested scenes to statistically compare inter-scene generalizability, as detailed in the \emph{results} paragraph below.
To acquire enough data for robust statistical distributions, we included $72$ repeats per condition on each scene, with fully randomized order.
Therefore, the experiment generated $12$ participants $\times 3$ scenes $\times 3$ conditions $\times 72$ repeats $=7776$ trials in total.
We avoided participant fatigue by partitioning the study into $6$ blocks, with each block containing trials from only one scene.
Additionally, the scene order was fully counterbalanced with a Latin square to avoid carry-on effects.

\paragraph{Results}
The second row of \Cref{fig:results:study} summarizes the results (see \Cref{sec:supp:study:fullStats} for the full visualization). 
To measure the model's applicability and generalizability,
we compare its predictions with the obtained human data along multiple axes, including unseen conditions (\Cref{fig:results:study:fit:all}), participants (\Cref{fig:results:study:fit:user}), and scenes.
Specifically,
\begin{enumerate}[leftmargin=*]
\item Across the $3$ conditions, $\conditionEvalStudyCombinedShort$ exhibits the fastest average offset time ($.49\pm .16$), compared to $\conditionEvalStudyVergent$ ($.58\pm .13$) and $\conditionEvalStudyCombinedLong$ ($.52\pm .13$) conditions.
The trend agrees with the model's prediction for $\conditionEvalStudyCombinedShort$/$\conditionEvalStudyVergent$/$\conditionEvalStudyCombinedLong$, as $.44\pm.13/.60\pm.15/.54\pm.16$.
The predictions for $\conditionEvalStudyVergent$ in \Cref{fig:results:study:fit:all} appear to be slightly higher than measured data, however, 
K.S. tests failed to reject the null hypothesis that the model prediction and the user-exhibited data are drawn from the same distribution ($p>.99$ for each condition).
A repeated measures ANOVA indicated that the condition had a significant effect on the offset time ($F_{2,22}=21.75, p<.001$). 

\item Across the $12$ participants, K.S. tests failed to reject the null hypothesis that the model prediction and the user-exhibited data are drawn from the same distribution ($p>.79$ for each).

\item Across the $3$ scenes, K.S. tests failed to reject the null hypothesis that the model prediction and the user-exhibited data are drawn from the same distribution ($p>.99$ for each scene).
A repeated measures ANOVA did not observe that the scene had a significant effect on the offset time ($F_{2,22}=1.93, p=.17$). 
We further calculated the \KLdiv{s} between observed data and model predictions for each scene to investigate whether the choice of scene affects model alignment.
The \KLdiv for archery/basketball/natural is $.52\pm.27/.56\pm.29/.54\pm.23$, respectively.
A repeated measures ANOVA did not observe that scene had a significant effect on the \KLdiv ($F_{2,22}=.51, p=.61$).
\end{enumerate}

\paragraph{Discussion}
The statistical analysis demonstrates the model's consistent capability of predicting and thus optimizing users' task performance during 3D visual target changes.
In addition to averaged offset times, the model also accurately predicts probability distributions with statistical accuracy, considering individual differences and sensory/behavioral randomness.
Our predictions are consistent with unseen conditions and participants, without being affected by novel and realistic scenes.
We also re-observe the remarkable fact that offset time performance is not positively correlated to the travel distance, again evidenced by a significant ``U-shape'' effect.

\section{Application Case Studies}
\label{sec:applications}

We apply our model to two applications considering 3D gaze movements. 
First, we explore how gaze movement variability between VR games can influence video game difficulty experienced by players.
Second, we make recommendations for scene-aware design and placement of 3D UI elements to minimize the cost of users' target changing in scenarios such as automotive head-up displays (HUD).

\subsection{Gaze Movement Performance in Games for VR vs. 2D}
The relationship between human performance in video games and target placement has been studied in traditional 2D displays \cite{Duinkharjav:2022:IFI,kim2022display}.
In this case study, we consider whether the game-dependent content depth has an effect on this performance.
Since gaming in 2D does not involve vergence movements, our evidence in \Cref{sec:pilot} suggests that gaze movements would be faster than in 3D environments.
To measure the scale of this difference across display environments as well as individual games, we conduct a numerical simulation using our model.

\paragraph{Setup}
We experiment with a large-scale VR player behavior dataset established by \citet{aizenman2022statistics}.
The dataset investigates how often users fixate at various depths during gameplay.
It contains games which mimic four top-rated games on Steam\footnote{\url{https://store.steampowered.com/vr/\#p=0\&tab=TopSellers}}:
\emph{Job Simulator}$^\text\textregistered$,
\emph{Arizona Sunshine}$^\text\textregistered$,
\emph{Beat Saber}$^\text\textregistered$, and
\emph{Pistol Whip}$^\text\textregistered$.
With this data, we can simulate various gaze shifts between fixations $h_{\textit{f(ixation)}}$ that occur during real gameplay and use our model to predict the corresponding average offset time.
Concretely, the distribution of gaze fixation depth is described via a probability density function,
$h_f (\vergentAng \mid G)$.
The PDF value at some vergence angle, $\vergentAng$, represents the proportion of total time spent fixating at that depth when a user plays a given game $G$.

We model each gaze movement during play as originating and targeting two fixation points sampled from the same distribution $h_\textit{f}$.
Given an origin and target vergence angles, $\vergentAng^o$ and  $\vergentAng^t$, the joint probability density, $h_\textit{m(ovement)}(\vergentAmp)$, is equal to
\begin{align}
h_\textit{m}(\vergentAmp = \vergentAng^t - \vergentAng^o \mid G) =
h_\textit{f}(\vergentAng^t \mid G) \times h_\textit{f}(\vergentAng^o \mid G).
\end{align}
Using this distribution of vergence movement amplitudes, $h_\textit{m}$, as a weight factor, we compute the mean gaze movement offset times at all saccade amplitudes our model supports (i.e., $\saccadeAmp \in [4^\circ, 12^\circ]$).

\begin{figure}
    \centering
    \includegraphics[width=0.96\linewidth]{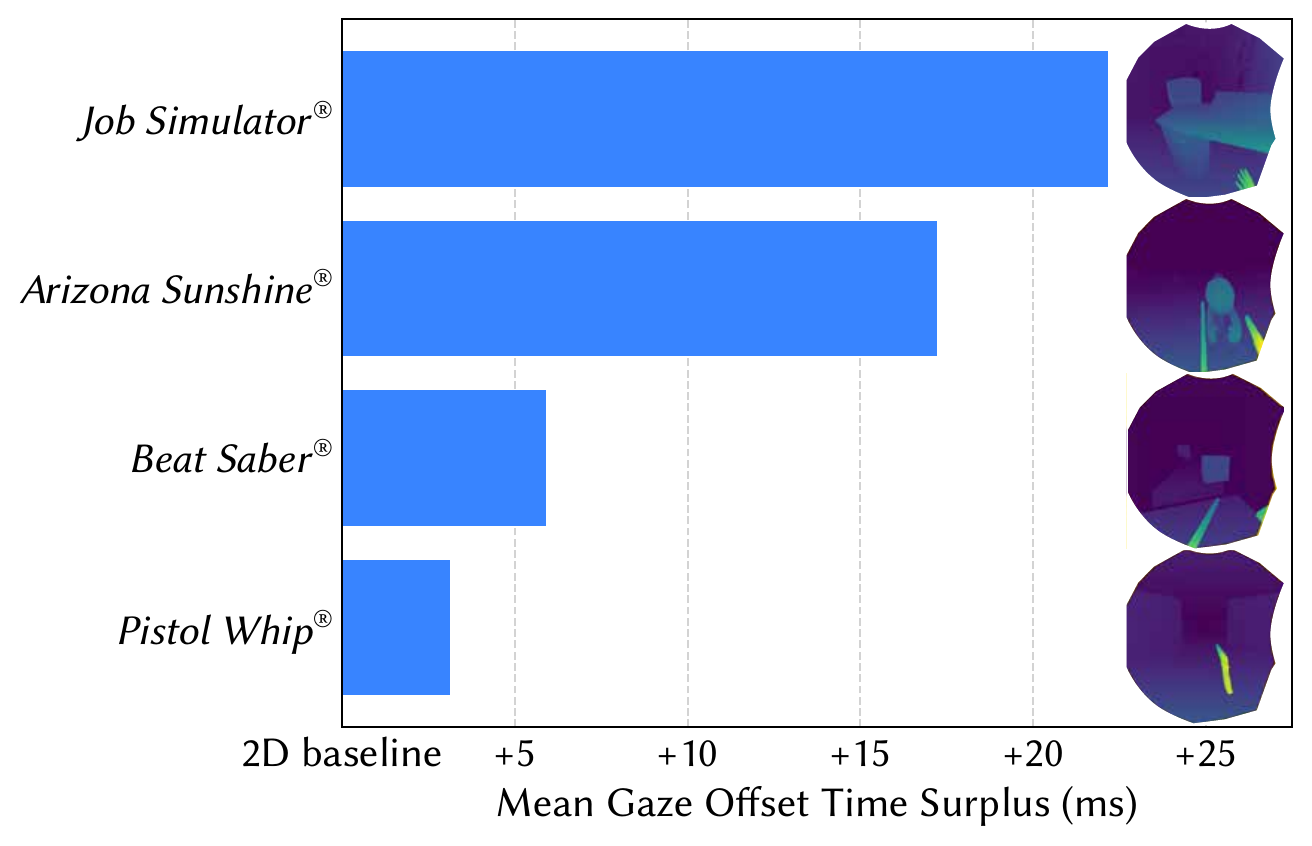}
    \Caption{Measuring target-shifting offset times in VR games.}{%
        Variability in the depth of salient regions in VR games induces longer gaze movement offset times due to combined vergence-saccade gaze movements.
        Representative depth-buffer frames from each image are shown as insets for each game.
        Games with higher variation in depth (\emph{Job Simulator}$^\text\textregistered$ and \emph{Arizona Sunshine}$^\text\textregistered$) exhibit longer offset times as predicted by our model.
        Traditional 2D video games do not involve depth changes during gaze movements, and therefore have a faster average offset time of $354\ \text{ms}$, shown here as a ``baseline'' for comparison.
    }
    \label{fig:application:gameperf}
\end{figure}

\paragraph{Results and discussion}
We visualize our main results in \Cref{fig:application:gameperf}.
Across all gaze depths reported by Aizenman~et~al.~\shortcite{aizenman2022statistics}, $98.7\%$ of the duration was fixated at vergence angles $\vergentAng \leq 8.4^\circ$ --- the maximum supported by our model.
In analysis, we excluded the remaining $1.3\%$ data.
The baseline 2D condition without vergence movements between fixations (i.e., $\vergentAmp=0$) exhibits the fastest offset times of $354\ \text{ms}$.
The mean offset times for the four games are, on average, $10\ \text{ms}$ slower compared to the baseline 2D condition.
\emph{Job Simulator}$^\text\textregistered$ and \emph{Arizona Sunshine}$^\text\textregistered$ present a mean gaze offset time of around $20\ \text{ms}$ more than baseline, while \emph{Beat Saber}$^\text\textregistered$, and \emph{Pistol Whip}$^\text\textregistered$ present a mean gaze offset time of around $5\ \text{ms}$.

The additional time and effort resulting from stereoscopic eye movements in different games will likely translate to increased difficulty.
Notably, the performance regression varies across games and depends on the scale of players' gaze depth variance.
These results suggest that gaming in VR comes with a ``performance overhead'' when compared to playing in 2D.
Games that feature more salient objects at shallow depths such as \emph{Job Simulator}$^\text\textregistered$ and \emph{Arizona Sunshine}$^\text\textregistered$ result in up to $20\ \text{ms}$ longer gaze offset times compared to the other two games where very little performance is lost.
Further investigations to characterize the relationship between gaze offset times and player-experienced difficulties are interesting future work but beyond the scope of this research.

\subsection{Scene-Aware Optimization for 3D User Interface}

\begin{figure*}[t]
    \centering
    \subfloat[observer view]{
        \includegraphics[width=0.255\linewidth,valign=t]{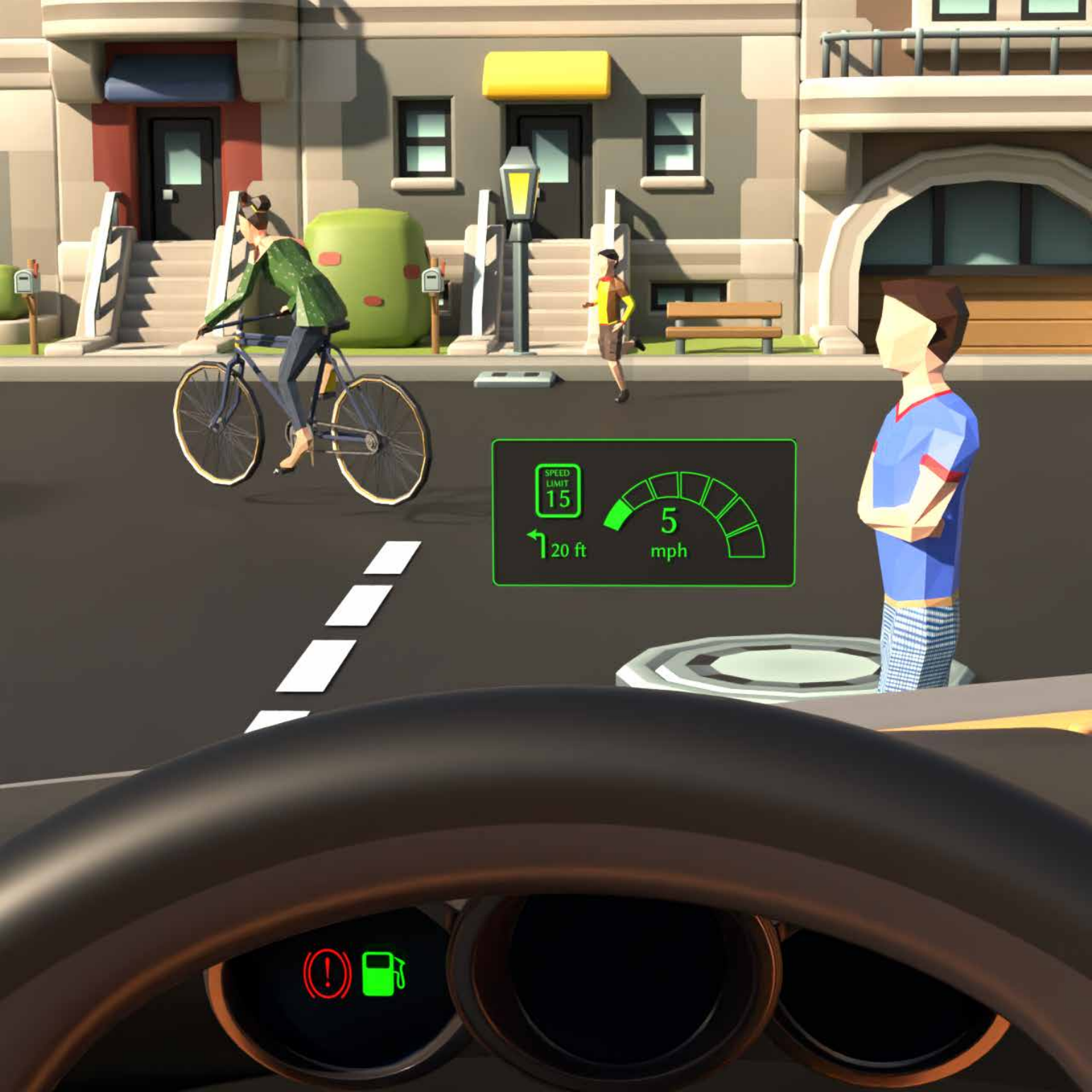}
        \label{fig:application:heatmap-visualization:scene}
    }%
    \hfill
    \subfloat[annotations]{
        \includegraphics[width=0.255\linewidth,valign=t]{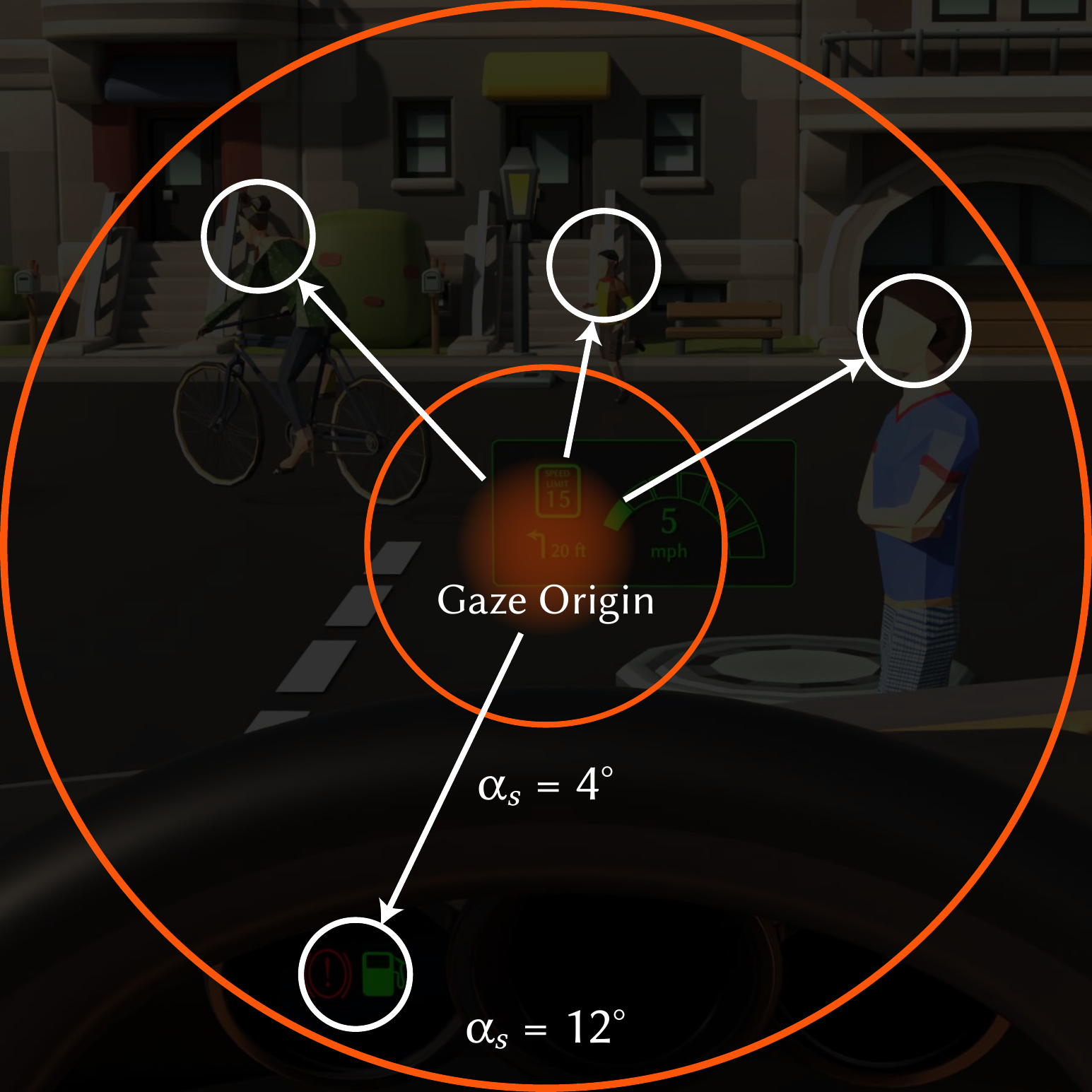}
        \label{fig:application:heatmap-visualization:annotations}
    }%
    \hfill
    \subfloat[gaze movement mean offset predictions]{
        \includegraphics[width=0.44\linewidth,valign=t]{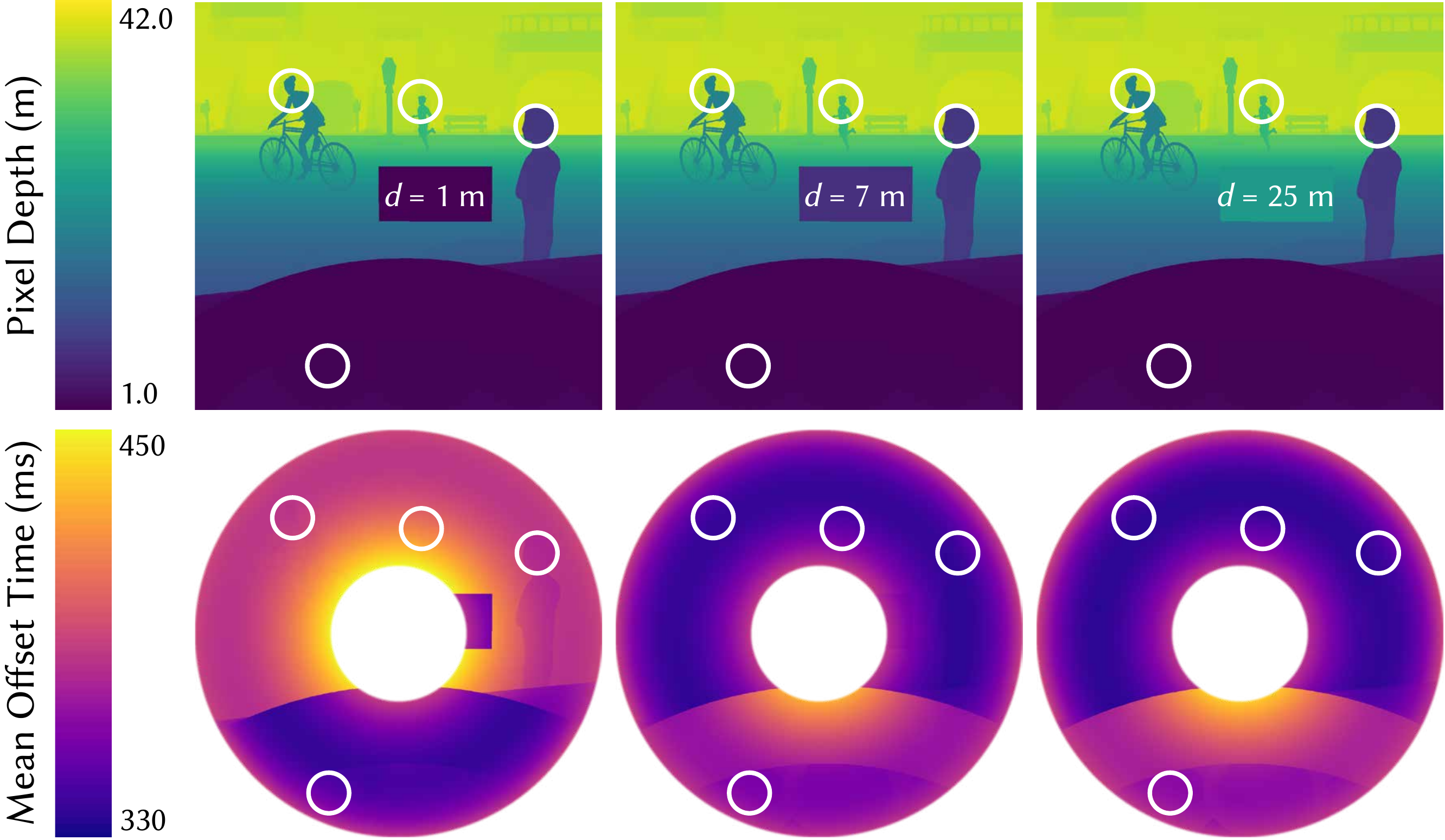}
        \label{fig:application:heatmap-visualization:heatmaps}
    }%
    \Caption{Predicted gaze movement offset times with vehicle HUD projected at various depths.} {%
The offset time varies when a driver shifts their gaze from the green  HUD virtual dashboard \subref{fig:application:heatmap-visualization:scene} to different peripheral targets \subref{fig:application:heatmap-visualization:annotations}, depending on the depth discrepancy between the source and target depths.
\subref{fig:application:heatmap-visualization:heatmaps} If the gaze origin is placed at the same depth as the car interior ($d \approx 1\ \text{m}$), gaze movements towards these locations are faster ($346\ \text{ms}$ at $1\ \text{m}$ compared to $359/365\ \text{ms}$ at $7/25\ \text{m}$).
In other words, as the depth of the gaze origin moves further ($d \approx 25\ \text{m}$), the gaze offset towards the car interior begins to increase.
However, for the goal of minimizing the offset time required to change gaze to the pedestrian on the right, a medium depth of $d \approx 7\ \text{m}$ is optimal ($342\ \text{ms}$ at $7\ \text{m}$ compared to $376/343\ \text{ms}$ at $1/25\ \text{m}$).
    } 
\label{fig:application:heatmap-visualization}
\end{figure*}

The surging automotive head-up displays (HUD) and wearable AR devices raise new demands in user-centric 3D interface design.
Suboptimal designs may slow users' reactions and cause dangers \cite{sabelman2015real}.
When it comes to HUD interface, a desirable design target is the ``optimal'' virtual projection distance that preserves or even accelerates drivers' reaction  to road conditions (see \Cref{fig:application:heatmap-visualization:scene}), in addition to factors such as focal depths. 
However, the optimization still remains debated and thus confounds designs.
For example, while some literature suggests the distance to be $2.5-4\ \text{m}$ \cite{betancur2011physical}, some manufacturers instead designed it as $10\ \text{m}$%
\footnote{\url{https://media.mbusa.com/releases/release-9e110a76b364c518148b9c1ade19bc23-meet-the-s-class-digital-my-mbux-mercedes-benz-user-experience}}.
Our model provides a quantitative metric for drivers' target-reaching time as a consequence of varying HUD projection distances. 

Specifically, as annotated in \Cref{fig:application:heatmap-visualization:annotations}: if the driver were to initiate a gaze movement from looking at the HUD image, depending on the depth of the UI element as well as the target location, the gaze offset times would vary anywhere between $330 - 450\ \text{ms}$ (\Cref{fig:application:heatmap-visualization:heatmaps}).
Therefore, driving assistant applications could leverage the predictions in gaze offset to adjust the placement of UI elements, or to provide timely intervention/alerts in case of emergencies.
While the specific optimization goal for object placement will vary depending on the application, we conducted an example optimization using our model without loss of generality.
Specifically, we leverage large-scale datasets to collect the depth distribution of various scenes and suggest the ideal placement of a ``HUD overlay image'' which would minimize the average gaze offset time from the display element to arbitrary points of focus within the scene.

\begin{figure}[t]
    \centering
    \includegraphics[width=\linewidth]{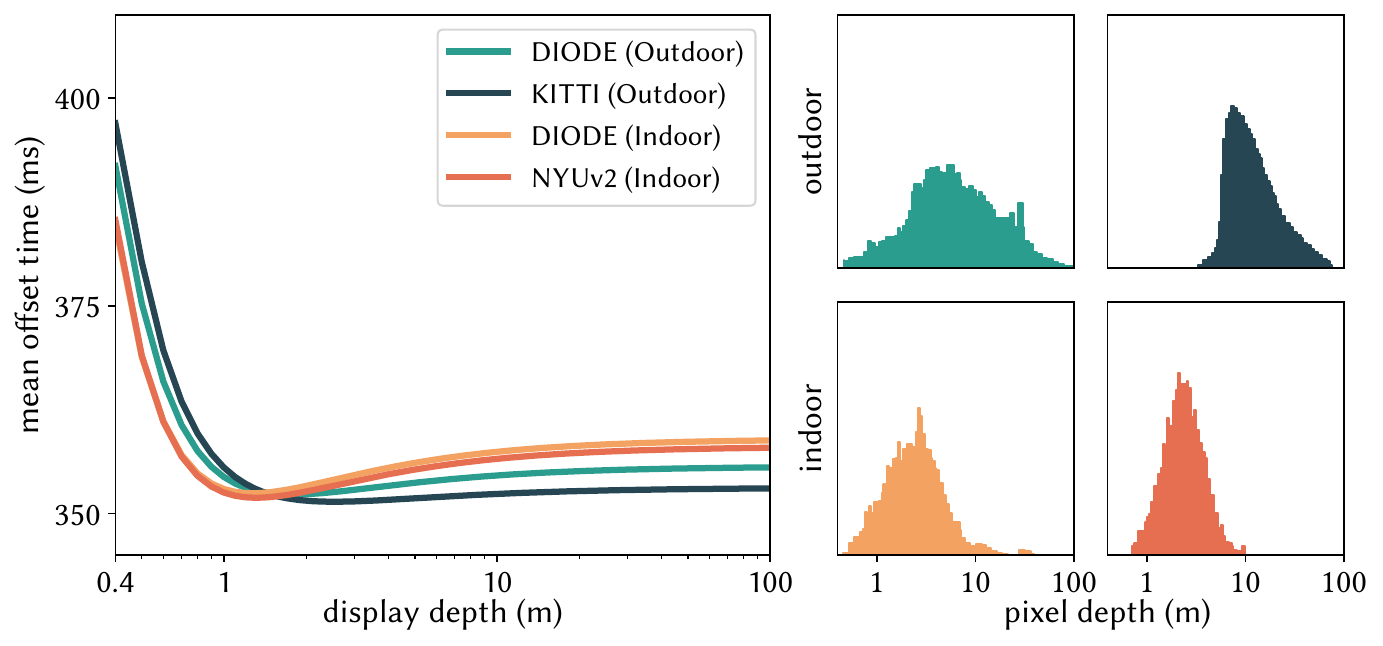}
    \Caption{Approximating offset times for VR/AR displays in natural scenes.}{%
    (left): By leveraging our model and a variety of large-scale datasets, we measure the average gaze movement offset time (Y-axis) originating from a HUD or AR display at various projection distances (X-axis) towards random locations in a natural 3D environment.
    We use publicly available datasets containing depth information in indoor and outdoor scenes.
    (right): shows the statistical density (Y-axis) of each dataset's per-pixel depths (X-axis).
    }
    \label{fig:application:depthtrend}
\end{figure}

\Cref{fig:application:depthtrend} shows our experimental results with two datasets containing depth maps of natural outdoor environments; DIODE \cite{vasiljevic2019diode} ($18,206$ frames), KITTI \cite{geiger2012we} ($12,919$ frames).
The average distances of objects are visualized in the top row of the histograms.
Assuming a starting gaze centered on a HUD overlay image, positioned at some depth, $d_{HUD}$, we measure the average gaze offset time, $\expectation[\travelTime]$, for saccade amplitudes uniformly sampled from $\saccadeAmp \in [4^\circ, 12^\circ]$, and depth targets sampled from the dataset depth histograms.
The resulting relationship between $d_{HUD}$ and $\expectation[\travelTime]$ is visualized in \Cref{fig:application:depthtrend}.
Due to the differentiable nature of our model, we can optimize $d_{HUD}$ to minimize $\expectation[\travelTime]$ via gradient descent.
As a result, the optimal image placements, $d_{HUD}^*$, are $1.8\ \text m$ and $2.5\ \text m$ for the outdoor DIODE and KITTI datasets.
Beyond HUD in outdoor environments, we may also leverage the model for AR devices in indoor scenarios. Therefore, we further leveraged the indoor portion from DIODE ($9,652$ frames), and NYUv2 \cite{silberman2012indoor} ($407,024$ indoor frames).
Intuitively, the depths that minimize $\expectation[t]$ are smaller for indoor datasets because more objects are closer in the distance.
Indeed, we found $1.3\ \text m$ to be the optimal projection depths for both the indoor-DIODE and NYUv2 datasets.

Our model helps design HUD displays in various applications, as the optimized image placements clearly vary significantly with scenes, e.g.\, indoor or outdoor ones.
They can also be further optimized by using distributions of saccade amplitudes that are more representative of each application.

\section{Limitations and Future Work}
\label{sec:future-work}

\paragraph{Initial depth and eccentricity}
Our combined vergence-saccade model measures the angular displacement in 3D without considering the initial fixation depth and eccentricity, even though both of these factors do influence eye movement offset time.
Specifically, prior literature suggests that convergence/divergence-only movements show a linear correlation for offset times \cite{templin2014modeling}, while off-axis movements that maintain focal depth are much more complex, and require consideration of both vertical/horizontal eccentricity and ocular-motor anatomics \cite{van2007sources}.
In order to develop a model that predicts gaze offset times between arbitrary points in 3D space, we would need to individually measure and account for all these factors as a high-dimensional grid of conditions.
Our main focus of this research is to demonstrate the importance and possibility of modeling gaze offset times for computer graphics applications; therefore, we plan to investigate all the factors above in future work.

\paragraph{Influence of accommodation and peripheral stereoacuity}
Vergence accommodation conflict may, in addition to discomfort, also cause incorrect visual fidelity \cite{march2022impact} and depth acuity \cite{sun2020eccentricity}, thus potentially degrading target localization accuracy. 
Similarly, the inherent mismatch between the geometric and empirical horopters may result in poor stereoacuity (and therefore localization) for targets at farther eccentricities along the iso-vergence circle \cite{ogle1952limits}. 
Additionally, accommodation speeds have been shown to be slower than vergence speeds \cite{heron2001dynamics}; hence, while our methods have comprehensive predictive capability in VR and pass-through AR devices (such as the Oculus Quest, and Apple Vision Pro), future investigations are necessary to fully model the latency of accommodation in \emph{see-through} AR devices.
Our stimuli cover a conservative range of vergence depths and eccentricities, with targets placed close to where the geometric and empirical horopters meet, and having little to no VAC.
While this range is appropriate for the contemporary (vergence-only) VR/AR displays \cite{aizenman2022statistics}, however, future work on understanding and optimizing for the influence of accommodation on 3D temporal visual behaviors may shed light on new performance-aware metrics to guide 3D display optics design.

\paragraph{Reaction time and image-space features}
Throughout this paper, we eliminated, as much as possible, any image-dependent variance in reaction time. Therefore, our measured offset time is primarily influenced by biomechanical responses to the spatial distribution of the stimuli, and not influenced by task difficulties or image characteristics such as contrast and spatial frequency \cite{devillez2020bimodality,lisi2019gain}. 
Exploring the combined effect of cognitive load or image characteristics on reaction time may add new building blocks for comprehensive measurements of visual performance.

\paragraph{Eye-head coordination}
During free-viewing, head movements often accompany eye movements and we tend to rotate our heads toward visual targets, especially for large eccentricities beyond $15^\circ$ \cite{bahill1975most}.
Our model does not predict the duration or impact of this concurrent head movement.
However, even though moving the head to center the target is a slower movement that typically completes after initial eye movement \cite{sauglam2011optimal}, our retinal image during the re-centering phase is stabilized, similar to Vestibular Ocular Reflex.
Hence, our model's predictions are likely to continue to be useful as they identify the earliest point after initial eye movement at which the target is clearly visible.
We hope that future work in eye-head movement validates this expectation.

\section{Conclusion}
\label{sec:conclusion}
We statistically measure and model the correlation between visual target displacement in 3D and eye movement offset time. 
Our data and model reveal a remarkable fact about eye movements in the 3D world: although combining a saccadic movement with a vergence movement accelerates motion towards a target in depth, the acceleration effect shows a surprisingly non-monotonic U-shape effect.
Moreover, the model accurately predicts absolute temporal performance on this task without individual normalization. This is primarily because offset time for eye movements is mainly a biophysical phenomenon and not a cognitive one.
We hope the research presented here inspires a new frontier exploring exciting questions about eye movements in 3D. 
For example, what contributes to variation in our target acquisition speeds? 
How do the surging virtual layers added to the physical world influence our visual attention shifts, and thus safety? 
And finally, how can we build future virtual environments that boost human performance in taking actions, even to outperform ourselves in the physical world? 
%\clearpage

% Uncomment acknowledgement section only for camera-ready
\begin{acks}
We would like to thank Avigael Aizenman and Agostino Gibaldi for insightful advice on processing stereo gaze data, and support in leveraging the video game gaze behavior data in their work \shortcite{aizenman2022statistics}.
This project is partially supported by the National Science Foundation grants \#2225861 and \#2232817, and a DARPA PTG program.
\end{acks}

% Bibliography
%%% -*-BibTeX-*-
%%% Do NOT edit. File created by BibTeX with style
%%% ACM-Reference-Format-Journals [18-Jan-2012].

% Supplementary material
\appendixpageoff
\appendixtitleoff
\renewcommand{\appendixtocname}{Supplementary material}
\begin{appendices}
\crefalias{section}{supp}
\normalsize

\include{\jobname-support}
\end{appendices}

\end{document}